\documentclass[12pt]{article}

\def\ref#1{\noindent \parshape=2 0pt 400pt 20pt 380pt #1}

\begin{document}

\centerline{\Large Electron acceleration due to high frequency}
 
\centerline{\Large instabilities at supernova remnant shocks}

\vskip 0.5cm

\centerline{\large M. E. Dieckmann,$^{\rm a,}$\footnote{Present address:
Institutet f\"or teknik och naturvetenskap, Link\"opings Universitet, Campus 
Norrk\"oping, S--601 74 Norrk\"oping, Sweden} K. G. 
McClements,$^{\rm b,}$\footnote{Corresponding author. E--mail: 
k.g.mcclements@ukaea.org.co.uk} S. C. Chapman,$^{\rm a}$}

\centerline{\large R. O. Dendy,$^{\rm b,a}$ and L. O'C. Drury$^{\rm c}$}

\vskip 0.3cm

\centerline{$^{\rm a}$ Space and Astrophysics Group, Department of Physics, 
University of Warwick,} 

\centerline{Coventry, CV4 7AL, UK}  

\vskip 0.3cm
              
\centerline{$^{\rm b}$ EURATOM/UKAEA Fusion Association, Culham Science 
Centre, Abingdon,} 

\centerline{Oxfordshire OX14 3DB, UK}
 
\vskip 0.3cm

\centerline{$^{\rm c}$ Dublin Institute for Advanced Studies, 5 Merrion Square,
Dublin 2, Ireland}

\vskip 0.3cm

\centerline{To appear in {\it Astronomy and Astrophysics} (accepted: February 
15 2000)}  

   \begin{abstract}

   Observations of synchrotron radiation across a wide range of
   wavelengths provide clear evidence that electrons are accelerated to
   relativistic energies in supernova remnants (SNRs). However, a viable
   mechanism for the pre--acceleration of such electrons to mildly 
   relativistic energies has not yet been
   established. In this paper an electromagnetic particle--in--cell 
   (PIC) code is used to simulate acceleration of electrons from
   background energies to tens of keV 
   at perpendicular collisionless shocks associated
   with SNRs. Free energy for electron energization is provided by ions 
   reflected from the shock front, with speeds greater 
   than the upstream electron thermal speed. The PIC simulation
   results contain several new features, including:
   the acceleration, rather than heating, of electrons via the Buneman
   instability; the acceleration of electrons to speeds exceeding those
   of the shock--reflected ions producing the instability;
   and strong acceleration of electrons perpendicular to the
   magnetic field. Electron energization takes place through a
   variety of resonant and non--resonant processes, of which the
   strongest involves stochastic wave--particle interactions. In SNRs
   the diffusive shock process could then
   supply the final step required for the production
   of fully relativistic electrons. The mechanisms identified in this
   paper thus provide a possible solution to the electron
   pre--acceleration problem.

   \end{abstract}

\section{Introduction}

The detection of radio synchrotron emission from shell--type supernova
remnants (SNRs) is a clear indication that electrons of typically GeV
energy are being accelerated in such objects. There is now convincing
evidence that synchrotron emission from some remnants extends to
X--ray wavelengths (Pohl \& Esposito 1998): this implies the presence
of electrons with energies of order 10$^{14}$eV. The prime example of
this is the remnant of SN1006, recent observations of which using the
ASCA (Koyama et al. 1995) and ROSAT (Willingale et al. 1996)
spacecraft show that X--ray emission from the bright rim has a hard,
approximately power--law spectrum. In contrast, emission from the
centre is softer, with a strong atomic line component. The sharp edges and
strong limb brightening observed at both X--ray and radio wavelengths
indicate that: the acceleration site is the strong outer
shock bounding the remnant; the acceleration is continuous; and
the local diffusion coefficient of electrons near the shock front
is substantially reduced relative to that in the general interstellar
medium (Achterberg et al. 1994). The possibility that the X--ray
emission from SN1006 is thermal bremsstrahlung has been
examined by Laming (1998), and found to be less tenable than the
synchrotron interpretation.           

There is thus extensive observational evidence that the strong
collisionless shocks bounding shell--type SNRs accelerate electrons to
relativistic energies. The standard interpretation of extragalactic
radio jet observations is also based on the premise that the
relativistic electrons responsible for observed synchrotron emission
are produced by shocks, although in this case the shock parameters are
much less certain. Heliospheric shocks, on the other hand, do not
generally appear to be associated with strong electron acceleration,
perhaps because the Mach numbers of such shocks are much lower than
those of SNRs and extragalactic radio jets, although Anderson et al. (1979)
have published data showing that keV electrons 
are produced in the vicinity of the perpendicular bow shock 
of the Earth.

While diffusive shock acceleration (Axford et al. 1977; Krymsky 1977; Bell 
1978; Blandford \& Ostriker 1978) provides an efficient 
means of generating highly energetic electrons from an already mildly 
relativistic threshold, and can operate at oblique shocks as well as 
parallel ones (Kirk \& Heavens 1989),
the ``injection'' or ``pre--acceleration'' question 
remains very open: by what mechanisms can electrons be accelerated from 
background (sub--relativistic) energies to mildly relativistic energies
(Levinson 1996)? In 
this paper, we investigate one of the possible answers, which has attractive 
``bootstrap'' characteristics. Specifically, we suggest that waves are excited
by collective instability of the non--Maxwellian population of ions reflected 
from a perpendicular shock front, and that these waves damp on
thermal electrons, thereby accelerating them. Such a process was first
proposed as a candidate acceleration mechanism for cosmic ray
electrons by Galeev (1984). This model was developed
further by Galeev et al. (1995), with the added ingredient
of macroscopic electric fields implied by the need to maintain
quasi--neutrality in a plasma with an escaping population of
electrons. McClements et al. (1997) carried out a
primarily analytical study of electron acceleration by ion--excited
waves at quasi--perpendicular shocks, which was necessarily restricted to
quantifying linear r\'egimes of wave excitation and particle
acceleration, in relation to inferred shock parameters. 

Instabilities driven by shock--reflected ions at
SNR shocks have also been invoked by
Papadopoulos (1988) and Cargill \& Papadopoulos
(1988) as mechanisms for electron heating, rather than
electron acceleration. On the basis of a simple analytical
calculation, Papadopoulos predicted that strong electron
heating would occur at quasi--perpendicular shocks with ``superhigh''
Mach numbers (specifically, shocks with fast magnetoacoustic Mach numbers $M_F
> 30$--40) through the combined effects of Buneman
(two--stream) and ion acoustic instabilities. In this model the
Buneman instability, driven by the relative streaming of
shock--reflected ions and upstream electrons, heats the electrons to a
temperature $T_e$ much greater than the ion temperature $T_i$: in
these circumstances, the ion acoustic instability can be driven
unstable if there is a supersonic streaming between the electrons and
either reflected or non--reflected (``background'') ions. Using a hybrid
code, in which ions were treated as particles and electrons as a
massless fluid, Cargill \& Papadopoulos (1988) found that
the electron heating predicted by Papadopoulos (1988)
could occur in a self--consistently computed shock structure. However,
as Cargill \& Papadopoulos point out in the last paragraph of their
1988 paper, the use of a fluid model for the electrons means that
hybrid codes cannot be used to investigate electron
acceleration. Recently, Bessho \& Ohsawa (1999) have used a
particle--in--cell (PIC) code to investigate acceleration of electrons
from tens of keV to highly relativistic energies at oblique shocks in 
which the electron gyrofrequency $\Omega_e$ exceeds the electron plasma
frequency $\omega_{pe}$.     

An improved theoretical understanding of electron acceleration at shocks
is desirable not only for intrinsic interest, but also to enable  
observations of synchrotron and inverse Compton emission to be
related quantitatively to shock parameters. However, almost all work on
particle acceleration has concentrated on ions. There are several
reasons for this. First, upstream momentum and energy fluxes are dominated by
ions, and the shock structure problem therefore reduces essentially to
that of isotropizing the ion distribution. Second, much of our
understanding is based on the use of hybrid codes, in which electrons
are represented as a fluid: such codes cannot provide information on
electron acceleration. However, the very fact that electron dynamics does not
appear to be important for shock structure allows us to separate the
two problems: prescribing the ion parameters using the results of
hybrid code simulations, we can examine in detail physical processes
occurring on electron timescales. This is the approach followed in
this paper. We describe the results of a fully nonlinear 
investigation, carried out by large scale numerical simulation using a 
PIC code and backed up by analytical and numerical studies, 
of the underlying plasma physics mechanisms. We consider the case
$\omega_{pe} > \Omega_e$, which is qualitatively distinct from the
strongly--magnetized r\'egime investigated by Bessho \& Ohsawa (1999). 
Our primary goal is finding a mechanism capable of producing mildly
relativistic electrons: once they have attained rigidities comparable to
those of shock--heated protons, they can undergo resonant scattering,
and subsequent acceleration to relativistic energies can then proceed
via the diffusive shock mechanism. Our approach enables us
to test earlier predictions of both electron acceleration (Galeev
1984; Galeev et al. 1995; McClements et al. 1997) and electron heating 
(Papadopoulos 1988; 
Papadoulos \& Cargill 1988) at very high Mach number astrophysical
shocks. Simulation results are presented for a range of reflected ion
speeds in Sect. 2; plasma instabilities occurring in the
simulations, and other processes likely to play a role in electron
acceleration and heating at SNR shocks, are identified
in Sect. 3; and the results of these investigations are discussed
in Sect. 4.           

\section{Particle--in--cell code simulations}

To investigate wave excitation and particle acceleration in the vicinity of
a perpendicular SNR shock we use an electromagnetic 
relativistic PIC code described by Devine (1995). The
particle--in--cell principle (Denavit \& Kruer 1980) relies
on self--consistent evolution of electromagnetic fields and macroparticles
in sequential stages. Relativistic 
electromagnetic PIC codes have been used previously to simulate acceleration 
processes in astrophysical plasmas (e.g. McClements et al. 
1993; Bessho \& Ohsawa 1999). A distinctive feature of the code used
in the present study is the fact that the energy density of
electromagnetic or electrostatic fluctuations can be readily
determined as a function of frequency $\omega$, wavevector $k$ or time $t$:
this greatly facilitates the identification of any wave modes excited
in a particular simulation.  

The code has one space dimension $(x)$ and three velocity dimensions 
$(v_x,v_y,v_z)$. To model a plasma containing shock--reflected proton
beams, we construct a simulation box with 350 grid cells in the
$x$--direction and with the local
magnetic field {\bf B} oriented in the $y$--direction. McClements et
al. (1997) pointed out that, at any given point in the
shock foot, there are in fact two proton beams, one propagating away from the
shock, the other towards it. For simplicity, we assume in our PIC
model that the two beams propagate with equal speeds $u_{b\perp}$ in
opposite directions perpendicular to the magnetic field, and that
both background ions and electrons have zero net drift: thus, the simulated
plasma has zero current. Strictly speaking, this is unrealistic,
since, in self--consistent models of perpendicular
shocks, the magnetic field magnitude has a finite gradient along the
shock normal direction, and a finite current is then required by
Amp\`ere's law (Woods 1969). We will discuss this
approximation in Sect. 4. The frame of reference in each simulation is
the upstream plasma frame: time evolution in the simulation can thus
be interpreted as spatial variation in the shock foot, with $t=0$ in
the simulation corresponding to the interface between the undisturbed
upstream plasma and the foot. The size of the foot $L_{\rm foot}$ lies
approximately in the range $(0.3 - 0.7)v_s/\Omega_i$, where $v_s$ is
the shock speed and $\Omega_i$ is the upstream ion gyrofrequency (McClements
et al. 1997). Thus, if the simulation is to be confined to the
foot, the duration of the simulation should be no greater than
$$ t_{\rm max} = {L_{\rm foot}\over v_s} \simeq (88 - 205){2\pi\over
\Omega_e}, \eqno (1) $$   
where $\Omega_e$ is the electron gyrofrequency (the true proton/electron mass
ratio, 1836, was used in the simulations).
The simulations reported here lasted for either 70 or 135 electron
cyclotron periods $2\pi/\Omega_e$.    

The proton beams were assumed to be initially Maxwellian with thermal
speed $\delta u_{\perp} = 3 \times 10^5\,$ms$^{-1}$ ($\delta u_{\perp}$ being
defined such that the equivalent temperature in energy units is 
$m_p\delta u_{\perp}^2$, where $m_p$ is the proton mass), and a
range of perpendicular drift speeds $u_{b\perp} = 3.25v_{e0}$, $3.5v_{e0}$,
$5v_{e0}$ and $6v_{e0}$, where $v_{e0}$  is the electron thermal speed,
defined in the same way as $\delta u_{\perp}$ and initially set equal
to $3.75 \times 10^6\,$ms$^{-1}$ (this corresponds to an electron
temperature $T_e \simeq 9.3 \times 10^5\,\hbox{K} \simeq
80\,\hbox{eV}$). The value chosen for the total beam number
density, $0.33n_e$, is consistent with the highest values of this
parameter found in hybrid simulations of quasi--perpendicular shocks with
Alv\'enic Mach numbers $M_A$ ranging up to about 60 (Quest
1986). Cargill \& Papadopoulos (1988) used
$M_A = 50$ in their hybrid simulation of an SNR shock (it
was computationally difficult to simulate shocks with higher $M_A$). 
The density of each beam $n_b$ is, accordingly, one sixth of the
electron density $n_e$, so that the background proton density $n_i$
required by charge balance is
$0.67n_e$ (the background proton thermal speed $v_i$ was
set equal to $1.9 \times 10^5\,$ms$^{-1}$). 
The electron plasma frequency $\omega_{pe}/2\pi$ and
gyrofrequency $\Omega_e/2\pi$ in our simulations were set equal to $10^5$Hz and
$10^4$Hz, respectively, corresponding to $n_e \simeq 1.2 \times
10^8\,$m$^{-3}$ and magnetic field $B \simeq 3.6 \times 10^{-7}\,$T.
The ratio $\omega_{pe}/\Omega_e$ is typically of order $10^2$ or
higher in HII regions of the interstellar medium. We have
chosen a relatively low value of this ratio in order to study and
compare the effects of instabilities occurring on both the
$\omega_{pe}^{-1}$ and $\Omega_e^{-1}$ timescales.    

The electrons, background protons and each proton beam were represented,
respectively, by 3200, 800 and 7200 particles per cell. The use of a
relatively small number of background protons per cell is
justified by the fact that instabilities driven by the proton beams have much
higher frequencies than noise fluctuations associated with the
background protons: large numbers of electrons and beam protons in
each cell ensure a level of noise energy well below the wave energy
produced by the instabilities. In what follows we measure time in electron
cyclotron periods, using the notation $\tilde{t}=\Omega_e t/2\pi$. We
also define $\tilde{k}=kv_{e0}/\Omega_e$
(only waves propagating in the $x$--direction are represented), a 
normalized frequency $\tilde{\omega} = \omega/\Omega_e$, and 
$r = kv_{\perp}/\Omega_{e}$, $v_{\perp}$ being electron 
velocity perpendicular to the magnetic field.

In every simulation, transfer of energy from beam protons to electrons
was observed, but the power flux between the two species increased
dramatically when $u_{b\perp}$ was raised from $3.5v_{e0}$ to
$5v_{e0}$. Figure 1 is a time evolution plot of perpendicular kinetic
energy ${\cal E}_{\perp e} = \sum_j m_ev_{\perp j}^2/2$, where $m_e$
is electron mass and the summation is over all electrons in the
simulation box. Since the total electron number is constant, ${\cal
E}_{\perp e}$ can be regarded as a measure of the effective
perpendicular electron  
temperature (although it should be stressed at the outset that the
electrons do not always have a Maxwellian distribution). The energy is
normalized to its initial value, which was identical in the four simulations.
When $u_{b\perp} = 3.25v_{e0}$ and $3.5v_{e0}$ (upper plot)
the energy increases by approximately an order of magnitude in around
60--100 electron cyclotron periods; when $u_{b\perp} =
5v_{e0}$ and $6v_{e0}$ (lower plot) the energy increases by a factor of
about 40 within $\tilde{t} \simeq 15-30$. The perpendicular energies of
the other two particle populations, again normalized to the initial
electron energy, are
plotted versus time for the case of $u_{b\perp} = 6v_{e0}$ in
Fig. 2. In the case of the beam protons (upper plot), both bulk motion
energy and thermal energy are included. During the simulation the
beam proton energy drops by less than 1\%, while the background proton
energy (lower plot) rises by no more than about 10\% (in the other
simulations the perpendicular energies of the two ion species changed
by even smaller amounts). In absolute terms the energy gained
by background protons is very small compared to that lost by beam protons,
with almost all the energy being transferred to electrons: we will
demonstrate that the beam protons excite an instability which couples
them efficiently to electrons.   
 
In all the cases studied, electrons were energized in the direction 
perpendicular to the magnetic field. The upper plot in Fig. 3 shows, in
more detail than Fig. 1, the time evolution of ${\cal E}_{\perp e}$
(once again normalized to its initial
value) in the first 25 electron cyclotron periods of the simulation
with $u_{b\perp} = 6 v_{e0}$. The lower plot shows the time evolution
of $\langle \varepsilon_0
E_x(x,t)^2/2 \rangle$, where $\varepsilon_0$ is the permittivity of free
space, $E_x(x,t)$ is the $x$--component of the electric field, and the 
brackets $\langle \rangle$ denote a spatial average over the simulation box. 
In general, $E_x$ is the dominant field component:
since propagation in the $x$--direction only is represented, it 
follows that the waves excited are predominately electrostatic. Henceforth,
the term ``electric field'' refers to the $x$--component. The field
has a single value in each simulation box cell: the electrostatic
field energy density $\langle \varepsilon_0
E_x(x,t)^2/2 \rangle$ is calculated by summing $\varepsilon_0
E_x(x,t)^2/2$ over the box and dividing by the number of cells.
The energy density plotted in the lower frame of Fig. 3 is   
normalized to the perpendicular electron energy density at $\tilde{t} = 0$.
The electron energy grows rapidly in two main phases, at   
$\tilde{t} \simeq 5$ and $\tilde{t} \simeq 14$, and  
then continues to grow at a slower rate. The 
field energy is greatly enhanced at times when the particle kinetic 
energy is growing rapidly: this suggests strongly that the  
fields are involved in particle acceleration. In the case of the wave
energy burst at $\tilde{t} \simeq 5$, the field energy grows to a
level comparable to the electron kinetic energy at that time. The
energy of the burst occurring at $\tilde{t} \simeq 14$, on the other
hand, is much lower than that of the electrons.
Figure 4 shows the time evolution of ${\cal E}_{\perp e}$ and field
energy in the simulation with $u_{b\perp} = 3.25 v_{e0}$. The upper
plot shows ${\cal E}_{\perp e}$ growing on a timescale comparable to 
the transit time of the simulation box through
the shock foot. The lower plot shows that electrostatic field activity  
is again correlated with electron acceleration. Figure 4 resembles the
second of the two periods of wave growth in Fig. 3 (at
$\tilde{t} \simeq 14$), in that the wave energy is small compared to the
electron kinetic energy. 

We now consider the distribution of wave amplitudes in wavenumber
space. Figure 5 shows the time evolution of this distribution
in the simulation with $u_{b\perp} = 6 v_{e0}$. The grey scale
shows the base 10 logarithm of the wave amplitude 
obtained by Fourier transforming in space the electric field of 
one of two counter--propagating waves excited by the ion beams.   
The start of the burst
in wave energy in the lower plot of Fig. 3 at $\tilde{t} \simeq 3$ can
be identified with the burst at $\tilde{k} \simeq 1.8$ in Fig. 5. 
This reaches an amplitude of 35 Vm$^{-1}$, generating a harmonic at 
$\tilde{k} \simeq 3.6$. When the peak amplitude is reached there is an
increase in wave energy at $\tilde{k} < 1$. The frequency of this low 
$\tilde{k}$ noise is close to the upper hybrid frequency
$\omega_{uh} = (\omega_{pe}^2 +
\Omega_e^2)^{1/2}$. Its appearance correlates with the maximum
of the first wave burst at $\tilde{t} \simeq 5$ in the lower plot of
Fig. 3, and with the 
strong increase of electron kinetic energy in the upper plot,
suggesting that it arises from a redistribution of
wave energy and changes in the electron distribution. After $\tilde{t}
\simeq 8$, when the initial wave burst has disappeared, a more
broadband perturbation is generated at $\tilde{k} \simeq 1.3$, the
mean $\tilde{k}$ decreasing with time. At $\tilde{t}=14$ the wave
amplitude peaks at about 16$\,$Vm$^{-1}$: this is considerably lower
than the peak electric field of 35$\,$Vm$^{-1}$ in the first burst, but
nevertheless strong enough to generate two harmonics (at $\tilde{k}
\simeq 2.6$ and $\tilde{k} \simeq 3.9$).  

The corresponding plot for the simulation with $u_{b\perp} =
3.25v_{e0}$ is shown in Fig. 6. In this case instability occurs at
discrete, regularly--spaced values of $\tilde{k}$. Waves with
relatively high $\tilde{k}$ ($\simeq 4$) are the first to be driven
unstable: during the course of the simulation, the instability shifts
to lower discrete wavenumbers. Broadband noise develops   
at $\tilde{k} < 1$, as in Fig. 5, but at a later time in the
simulation ($\tilde{t} \simeq 35$). This appears to be associated with
a more gradual evolution of the electron distribution than that which
occurs in the simulation with $u_{b\perp}=6v_{e0}$. The difference in temporal
behaviour between Figs. 5 and 6 will be discussed later in this paper.
Figures 5 and 6 show that in both simulations the plasma eventually
stabilizes, on a timescale which depends on the beam velocity. 

The dependence of wave amplitude on $\tilde{k}$ and $\tilde{t}$ when
$u_{b\perp}=5v_{e0}$ is qualitatively similar to Fig. 5: after an  
intense burst early in the simulation, a wave with slowly--varying
amplitude is observed to cascade down in $\tilde{k}$ as time
progresses. The growth rate of the first wave burst is 20\% higher in
the simulation with $u_{b\perp} = 6v_{e0}$ than it is in the
simulation with $u_{b\perp} = 5v_{e0}$. In the former case, as
mentioned above, the peak amplitude of 
the second burst is 16$\,$Vm$^{-1}$, at 
$\tilde{k} = 1.25$ and $\tilde{t} = 14$; the corresponding figures
for the simulation with $u_{b\perp} = 5v_{e0}$ are 12$\,$Vm$^{-1}$,
$\tilde{k}=1.88$ and $\tilde{t} = 12$. The wave amplitude distribution in the
simulation with $u_{b\perp}=3.5v_{e0}$ is similar to Fig. 6: bursts of
wave activity occur at discrete $\tilde{k}$, with the high $\tilde{k}$
modes being driven unstable first. 

In principle, it is also possible to determine the time evolution of wave 
amplitude as a function of $\tilde{\omega}$ and $\tilde{k}$. However, in order
to obtain good frequency resolution it is necessary to average the
amplitude over times longer than the electron acceleration timescale.
Electrostatic waves in the electron cyclotron range propagating
perpendicular to the magnetic field include, for example, electron
Bernstein waves, whose dispersion relation depends on the electron
distribution. Since this is rapidly evolving, it can be difficult 
to interpret observed distributions of wave amplitude in
$\tilde{\omega}$ and $\tilde{k}$.   
However, we have found that the most strongly--growing waves in the
simulations invariably have $\tilde{\omega} \simeq
\tilde{k}u_{b\perp}/v_{e0}$: one can thus obtain the frequencies of
the high intensity modes in Figs. 5 and 6 by multiplying $\tilde{k}$
by $u_{b\perp}/v_{e0}$. By this means, it is straightforward to verify
that the modes excited early in both simulations have $\tilde{\omega}
\simeq 10$, and hence $\omega \simeq \omega_{pe}$. 

\section{Analysis of simulation results}

Short--lived bursts of narrowband wave activity, correlated with
rapid increases in electron kinetic energy, occur for all four values
of $u_{b\perp}/v_{e0}$ considered above.
These bursts appear throughout the simulations 
with $u_{b\perp} = 3.25 v_{e0}$ and $u_{b\perp} = 3.5 v_{e0}$; in the
case of $u_{b\perp} = 5v_{e0}$ and $u_{b\perp} = 6v_{e0}$, they appear
only at early times. In every case, the instability cascades to longer
wavelengths in the course of the simulation. In order to compare the simulation
results with those given by linear instability analysis (described in
the next subsection), we determine
growth rates for the first wave that interacts significantly with the
electrons: in such cases one may assume 
that the electrons are still represented by a single Maxwellian velocity 
distribution with thermal speed $v_{e0}$. 
The simulations provide the wavenumber $\tilde{k}$ of the unstable
wave modes and the electric field amplitude $E$. The real frequencies 
$\tilde{\omega}$ of the unstable wave modes are assumed to be equal to 
$\tilde{k}u_{b\perp}/v_{e0}$. The normalized growth rate 
$\gamma/\Omega_e$ is estimated by fitting an exponential to the plot
of wave amplitude versus time during the period of most rapid growth
in each simulation.  

The results of this analysis are shown in Table 1. The symbol $E_m$ denotes
the maximum value of $E$ during each simulation. In three of the four 
simulations there is a period of wave growth which can be described accurately
as exponential. In each case, the growth rate falls to zero, and
the wave decays: an example of this behaviour, for the case of
$u_{b\perp} = 6v_{e0}$, is shown in Fig. 7, where wave amplitude (defined 
as in Figs. 5 and 6) at $\tilde{k} = 1.8$ is plotted versus normalized time. 
A possible reason for wave collapse (observed in all four simulations) will be
discussed later in this paper. In the case of $u_{b\perp} = 3.25v_{e0}$, the 
mode referred to in Table 1 ($\tilde{k} \simeq 3.6$) is the second to be 
destabilized in that simulation. It appears to grow linearly 
rather than exponentially: for this reason, no figure is given for
its growth rate. The first mode to be destabilized in this simulation, 
with $\tilde{k} \simeq 3.9$, does not grow to a large amplitude (compared to
the noise level), and so it is difficult to determine its growth
rate. Later in the paper we will present evidence of wave--wave
interaction between the second mode excited ($\tilde{k} \simeq 3.6$) and the 
third mode excited ($\tilde{k} \simeq 3.3$), which may help to
explain the linear growth of the latter. 

\vskip 0.5cm

\noindent {\bf Table 1.} Parameters of highest intensity wave mode in
each simulation. 

\vskip 0.2cm

\noindent \begin{tabular}{lllll}

\hline

$u_{b\perp}/v_{e0}$ & $\tilde{k}$ & $\tilde{\omega}$ & $\gamma/\Omega_e$ & 
$E_m$ (Vm$^{-1}$) \\

\hline 

$6.0$ & 1.8 & 10.8 & 0.24 & 35 \\

$5.0$ & 2.15 & 10.7 & 0.2 & 23 \\

$3.5$ & 3.3 & 11.6 & 0.05 & 2.5 \\

$3.25$ & 3.6 & 11.7 & -- & 1.6 \\

\hline

\end{tabular}

\vskip 0.5cm

Let us now examine whether the growth rates derived from the PIC simulations
in Table 1 and Fig. 7 can be reproduced using linear
stability theory.       

\subsection{Linear stability analysis} 

The appropriate dispersion relation for electrostatic,
perpendicular--propagating waves with frequencies in the electron
cyclotron range and above, excited by an ion beam with a Maxwellian
distribution in $v_{\perp}$, is (Melrose 1986)   
$$ 1-{\omega_{pi}^2\over \omega^2} + 
{2\omega_{pb}^2\left[1+\zeta_b Z(\zeta_b)
      \right]\over k^2\delta u_{\perp}^2}
- {\omega_{pe}^2\over \omega}{e^{-\lambda_e}\over 
      \lambda_e}\sum_{\ell = -\infty}^{\infty}{\ell^2I_{\ell}\over 
      \omega - \ell\Omega_e} = 0 \,, \eqno (2) $$
where: $\omega_{pi}$, $\omega_{pb}$ are the background and beam ion plasma
frequencies; $Z$ is the plasma dispersion function, with argument
$\zeta_b \equiv (\omega-ku_{b\perp})/k\delta u_{\perp}$; and $I_{\ell}$ 
is the modified Bessel function of the first kind of order $\ell$ with
argument $\lambda_e \equiv T_ek^2/(m_e\Omega_e^2)$. Both species of
ion, having a much longer cyclotron period than the electrons, can be
treated as unmagnetized particles on the timescales of interest here.
Strictly speaking, there should be a term in Eq. (2) for each of the
two proton beams, but since they have mean perpendicular speeds
of opposite sign, and $\omega \simeq ku_{b\perp}$ is a prerequisite
for wave--particle interaction, we need only consider one of them.      

Solutions of Eq. (2) for complex $\omega$ in terms of real $k$ can be
readily obtained numerically, and compared with the simulation results
in Table 1. In Figs. 8 and 9 $\tilde{\gamma} \equiv
\rm{Im}(\tilde{\omega})$ is plotted versus $\tilde{k}$ for parameters
corresponding to the initial conditions of the simulations with
$u_{b\perp} = 6v_{e0}$ and $u_{b\perp} = 3.25v_{e0}$. In the former
case it can be seen that strong instability drive occurs at $\tilde{k}
\simeq 1.8$ with maximum growth rate $\tilde{\gamma} \simeq 0.25$, as
observed early in the simulation (Fig. 5 and Table 1). The unstable real
frequencies range from $\tilde{\omega} \simeq 8$ to $\tilde{\omega}
\simeq 10.8$, and are thus clustered around the dimensionless electron
plasma frequency ($\tilde{\omega} = 10$). The main instability appears to
be essentially unaffected by cyclotronic effects: the growth rate does
not depend on how close the frequency is to a cyclotron
harmonic. There are, however, two much weaker instabilities at $\tilde{k} <
1$, which are narrowband in both $\tilde{k}$ and $\tilde{\omega}$, the
real frequencies lying just below the second and third cyclotron
harmonics. In Fig. 9 instability occurs at $\tilde{k} \sim
3-4$, the corresponding real
frequencies again clustering around $\omega_{pe}$. In this case,
however, the instability is modulated by cyclotronic effects, as
in the simulation. Instability again occurs at $\tilde{k} < 1$, with
real frequency $\tilde{\omega} \simeq 1.8$.

The mode appearing early in the simulation with $u_{b\perp} = 6v_{e0}$
arises from a two--stream instability
(Buneman 1958). This can be driven by ions drifting relative to
electrons in an unmagnetized plasma: it can also occur in a
magnetized plasma, with ions drifting across the field, if
$\omega_{pe}/\Omega_e$ is sufficiently large, and the instability
drive is sufficiently strong. Electrons as well as ions are then
effectively unmagnetized and the appropriate dispersion relation is
(Melrose 1986)     
$$ 1 - {\omega_{pi}^2\over \omega^2} +
{2\omega_{pb}^2\left[1+\zeta Z(\zeta_b)
      \right]\over k^2\delta u_{\perp}^2}
 + {2\omega_{pe}^2\left[1+\zeta Z(\zeta_e)
      \right]\over k^2v_{e0}^2} = 0 \,, \eqno (3) $$ 
where $\zeta_e \equiv \omega/kv_{e0}$. In the frequency r\'egime of
interest here ($\omega \simeq \omega_{pe}$), it can be shown easily 
that the background ion term in Eq. (3) can be neglected. Letting the
thermal speeds of the two remaining species tend to zero, Eq. (3) reduces to
$$ 1-{\omega_{pb}^2\over (\omega-ku_{b\perp})^2}
      -{\omega_{pe}^2\over \omega^2} = 0 \,. \eqno (4) $$
This differs slightly from the original two--stream dispersion relation
analysed by Buneman (1958) in that the ions, rather
than the electrons, have a finite drift speed. Buneman's equation
becomes identical to Eq. (4) under the transformation $\omega
\rightarrow ku_{b\perp} - \omega$: using this, we can infer
from Buneman's analysis that Eq. (4) has a root $\omega = \omega_0
+i\gamma$, where real frequency $\omega_0$ and growth rate $\gamma$
are given approximately by       
$$ \omega_0 \simeq ku_{b\perp} -
\omega_{pb}^{2/3}\omega_{pe}^{1/3} \cos^{4/3}\theta, \eqno (5) $$ 
$$ \gamma \simeq \omega_{pb}^{2/3}\omega_{pe}^{1/3}
\cos^{1/3}\theta\sin \theta, \eqno (6) $$ 
$\theta$ being a parameter whose value depends on
$(ku_{b\perp} - \omega_{pe})/\omega_{pb}^{2/3}
\omega_{pe}^{1/3}$ (Buneman 1958): it is straightforward to
verify that the strongest wave growth occurs when $\theta = \pi/3$,
which corresponds to 
$ku_{b\perp} = \omega_{pe}$. If $ku_{b\perp} \gg 
\omega_{pb}^{2/3}\omega_{pe}^{1/3} \cos^{4/3}\theta$, it follows from
Eq. (5) that $\omega \simeq ku_{b\perp}$ and so the strongest
drive occurs at $\omega \simeq \omega_{pe}$. However, since $\theta$
can have a range of values, the instability has finite bandwidth, extending to
frequencies significantly below $\omega_{pe}$. Solving the full
Buneman dispersion relation [Eq. (4)] with $u_{b\perp} = 6v_{e0}$,
we obtain results which are almost identical to those
obtained from the magnetized dispersion relation [Eq. (2)], except, of
course, that the cyclotronic features at $\tilde{k} < 1$ in Fig. 8 do
not appear. Even in the case of $u_{b\perp} = 3.25v_{e0}$, Eq. (4)
yields instability at about the same wavenumbers and frequencies as
Eq. (2) [although the growth rates are somewhat higher in the case of
Eq. (4)]. The essential
difference between Figs. 8 and 9 is that the lower beam speed in the
latter yields lower growth rates: when $\tilde{\gamma}$ is
sufficiently small, the gyromotion of an electron in one wave growth
period cannot be neglected, and the instability is modified by
cyclotronic effects. However, the instability remains Buneman--like in
character.    

Further analysis of Eq. (2) indicates that the instability
growth rate is a slowly--decreasing function of $\delta
u_{\perp}/u_{b\perp}$: in the case of $u_{b\perp} = 6v_{e0}$, for
example, the maximum growth rate is around $0.06\Omega_e$ when $\delta
u_{\perp}/u_{b\perp} = 0.3$. The Buneman instability is thus robust,
in the sense that its occurrence is not critically dependent on the
velocity--space width of the reflected ion distribution. In any event,
the values of $\delta u_{\perp}/u_{b\perp}$ used in our PIC simulations
are broadly consistent with reflected beam ion distributions occurring
in the hybrid simulations of Cargill \& Papadopoulos (1988).          

The instabilities at $\tilde{k} < 1$ in Figs. 8 and 9 arise
from the interaction of a beam mode ($\omega \simeq ku_{b\perp}$) with
electron Bernstein modes. The existence of such instabilities can be
inferred analytically by taking the limit of Eq. (2) for cold beam and
background protons: 
$$1-{\omega_{pi}^2\over \omega^2} - {\omega_{pb}^2\over (\omega-ku_{b\perp})^2}
-{\omega_{pe}^2\over \omega}{e^{-\lambda_e}\over 
\lambda_e}\sum_{\ell = -\infty}^{\infty}{\ell^2I_{\ell}\over 
\omega - \ell\Omega_e} = 0 \,. \eqno (7) $$
In the absence of the proton beam term, Bernstein mode solutions of
Eq. (7) have frequencies which approach 
$\ell\Omega_e$ ($\ell = 1,2,3,...$) as $k \to \infty$ and
$(\ell+1)\Omega_e$ as $k \to 0$ (the long wavelength limit
is different for frequencies equal to or greater than the upper hybrid
frequency $\omega_{uh}$, which in the case of the simulations presented
in Sect. 2 is about 10$\Omega_e$). When $n_b \ne 0$, approximate
analytical solutions of Eq. (7) can be obtained by setting 
$\omega = ku_{b\perp} + i\gamma$ and solving perturbatively for
$\gamma$ in certain limits. For example, letting $\lambda_e \to 0$ and
assuming that $\omega$ does not lie close to a harmonic of $\Omega_e$, 
we obtain
$${\gamma\over \Omega_e} \simeq \left({m_e\over m_p}\right)^{1/2}
\left({n_b\over n_e}\right)^{1/2}\left({\omega^2\over \Omega_e^2} - 1
\right)^{1/2}\,. \eqno (8) $$   
Instability, corresponding to real $\gamma$, thus requires $\omega > \Omega_e$.
For $\omega$ 
sufficiently close to $\ell\Omega_e$, the electron term in Eq. (7) is 
dominated by the $\ell$--th harmonic, and instead of Eq. (8) we obtain   
$${\gamma\over \Omega_e} \simeq {\omega\over \Omega_e}\left({m_e\over 
m_i}\right)^{1/2}\left({n_b\over n_e}\right)^{1/2}{\lambda_e e^{\lambda_e}\over
\left(\ell I_{\ell}\right)^{1/2}}\left(1 - {\ell\Omega_e\over \omega}
\right)^{1/2}. \eqno (9) $$ 
Numerical solutions of Eq. (2) for $\omega \sim \Omega_e$ are broadly
consistent with Eqs. (8) and (9). In both these cases
the growth rate scales as $(m_e/m_p)^{1/2}$: in contrast, the Buneman
growth rate [Eq. (6)] scales as $(m_e/m_p)^{1/3}$. This helps to
explain the fact that in Figs. 8 and 9 the Buneman instability is
stronger than the lower frequency Bernstein instability.        
It should also be noted that most astrophysical plasmas have a higher
ratio of electron plasma frequency to gyrofrequency that that assumed
in the simulations ($\omega_{pe}/\Omega_e = 10$). Normalized to
$\Omega_e$, the Buneman growth rate scales as $\omega_{pe}/\Omega_e$,
and so the instability is less likely to be modified by
cyclotronic effects when $\omega_{pe}/\Omega_e > 10$. The electron
Bernstein modes exist because $T_e$ is finite: thus,
the transition from Buneman to electron Bernstein instability depends
on the value of $v_{e0}$. If the initial electron temperature in the
simulations had been lower than 80$\,$eV, the Buneman instability
would, again, have been affected to a lesser extent by cyclotronic
effects.       

\subsection{Nonlinear effects}

Figure 1 shows strong increases in electron kinetic energy
perpendicular to the magnetic field in all four simulations. There is
a strong correlation between acceleration and wave excitation via the
Buneman instability (Figs. 3 and 4). Although such waves can energize
electrons via Landau damping (Papadopoulos 1988), one
would expect this process to be of limited effectiveness when, as in
the present case, the
waves are propagating perpendicular to the magnetic field and have a
growth rate which is comparable to or less than $\Omega_e$.
It is likely therefore that the very strong electron acceleration observed in
the simulations is due at least in part to nonlinear processes. 

As noted previously, the second mode to be excited in the simulation
with $u_{b\perp}=3.25v_{e0}$ does not undergo an 
exponential growth phase. Figure 10 shows the time evolving wave 
amplitudes of this mode, at $\tilde{k} = 3.6$ (upper plot), and the third
mode to be excited, at $\tilde{k}=3.3$ (lower plot). The
amplitude at $\tilde{k}=3.6$ grows linearly up to $\tilde{t} \simeq
27$, and then collapses. The amplitude at $\tilde{k}=3.3$ grows exponentially
from $\tilde{t} \simeq 7$ to $\tilde{t} \simeq 14$, with
$\gamma/\Omega_e \simeq 0.04$: this is close to the
growth rate at $\tilde{k}=3.3$ found by linear stability analysis
(Fig. 9). The amplitude remains constant 
until $\tilde{t} \simeq 27$, and then grows linearly until 
$\tilde{t} \simeq  35$. The linear growth of the wave at $\tilde{k} =
3.3$ thus correlates strongly with the collapse of the wave at 
$\tilde{k} = 3.6$: this suggests wave--wave coupling.
The linear growth of the wave at $\tilde{k} = 3.6$ may, in
turn, be correlated with the decay of the first mode to be excited, at 
$\tilde{k} \simeq 3.9$
(see Fig. 6): the latter has the highest growth rate of waves in this
range, according to linear stability analysis (Fig. 9).

We now consider specific nonlinear processes which may be occurring
in the simulations. Karney (1978) examined the nonlinear
interaction of large amplitude electrostatic lower hybrid waves with
ions. The particle motion is described by a normalized Hamiltonian 
$$h = \frac{1}{2}{({p}_{x}+y)}^{2} + \frac{1}{2} {{p}_{y}}^{2}
-\alpha\sin{(y-\mu t)}, \eqno (10)$$
where: $\hat{\bf y}$, $\hat{\bf z}$ are, respectively, the wave
propagation and magnetic field directions; the canonical momentum
components $p_x$, $p_y$ are normalized to $m\Omega/k$, where $m$ and
$\Omega$ are particle mass and gyrofrequency; the position variable
$y$ is normalized to $1/k$; $\mu$ is wave frequency in units of
$\Omega$; and $\alpha$ is given by   
$$\alpha = \frac{E/B}{\Omega/k}, \eqno (11)$$ 
$E$ being the wave electric field amplitude. The first two terms in
the Hamiltonian describe the motion of the particle in the magnetic
field; the third term arises from the electrostatic wave. The system
can be regarded as consisting of two harmonic oscillators: one 
associated with the particle gyromotion around $B$, the other with the
wave. These oscillators are coupled by the parameter $\alpha$, the
value of which determines the extent to which the system phase space 
is regular or stochastic. Karney (1978) solved the Hamiltonian equations 
corresponding to Eq. (10) for a range of initial conditions, plotting 
normalized Larmor radius $r = kv_{\perp}/\Omega$ versus wave phase
angle $\phi$ at the particle's position, for successive transits of
the particle through a particular gyrophase angle. Particle
trajectories were thus represented as discrete sets of points in
$(r,\phi)$ space. For sufficiently small values of $\alpha$, all particles 
have regular orbits, represented by smooth curves in
$(r,\phi)$ space, spanning all values of $\phi$ and with only
small variations in $r$. When $\alpha$ exceeds a certain threshold
$\alpha_0$, ``islands'' appear in $(r,\phi)$ space, within which
particle trajectories are confined. Further increases in $\alpha$ lead
to the formation of more islands, which cause the phase space to become
stochastic: at sufficiently large $\alpha > \alpha_c$, the system
phase space is completely stochastic, with no regular orbits remaining. The
initial electron distributions in our simulations decrease  
monotonically in $v_{\perp}$: in such cases stochasticity in
phase space tends to favour particle diffusion to larger velocities,
i.e. acceleration.         

Karney (1978) obtained the following analytical estimate for $\alpha_0$:
$$\alpha_0 = \left | \frac{r(\omega/\Omega - \ell)}{\ell (d/dr) 
J_{\ell}(r)}  \right |, \eqno (12)$$
where $\ell\Omega$ is the cyclotron harmonic closest to $\omega$ and
$J_{\ell}$ is the Bessel function of order $\ell$.
Karney's analysis does not explicitly involve a 
particular type of wave or particle, or a specific particle
distribution function. The results can thus be applied to the case of
electrons interacting with electrostatic waves excited by the Buneman
instability, in which case $m = m_e$ and $\Omega=\Omega_e$.
Combining Eqs. (11) and (12) and using the identity
$$\frac{d}{dr}J_{\ell}(r) = \frac{\ell}{r}J_{\ell}(r) -
J_{\ell+1}(r), \eqno(13)$$ 
we infer that the critical electric field $E=E_i$ for island formation
in $(r,\phi)$ space is
$$E_i = \frac{v_{\perp}B_{0}|\mu -\ell|}{\ell|
\frac{\ell}{r}J_{\ell}(r) - J_{\ell+1}(r)|}. \eqno (14)$$
In general, it is not possible to determine analytically an expression
for the electric field amplitude $E=E_c$ corresponding to
$\alpha=\alpha_c$, above which the phase space becomes completely
stochastic. Karney obtained an empirical expression for $\alpha_c$,
based on numerical calculations with particular values of $\mu$ and
$r$, which may not be applicable to the simulation results discussed
here. However, island formation is a first step in the destruction of
regularity in the system phase space, and $E_i$ can be regarded as an
approximate threshold for stochasticity: electric field amplitudes
which are significantly higher than $E_i$ will convert regular
orbits at a particular $v_{\perp}$ into stochastic ones.
 
In the cases $u_{b\perp} = 5v_{e0}$ and $6v_{e0}$, linear
stability analysis indicates that wave growth occurs across a range of
frequencies $\omega \sim \omega_{pe}$, which includes cyclotron
harmonics: in such cases $\mu = \ell$, and any non--zero wave amplitude
$E$ will cause islands to be formed. In the case of the lower two beam
speeds, the unstable frequencies lie between cyclotron harmonics, and $E_i$ is
thus always finite. Table 2 lists the values of $E_i$ derived from
Eq. (14) that are required for comparison with the highest
intensity wave mode excited in each simulation. The actual peak
electric fields of these waves are given in Table 1.    

\vskip 4.0cm

\noindent {\bf Table 2.} Values calculated for $E_i$ using the 
wave parameters given in Table 1.

\vskip 0.2cm

\noindent \begin{tabular}{llllll}

\hline 

$v_{\perp}/v_{e0}$ & closest $\ell$ & $(\mu-\ell)$ & $kv_{\perp}/\Omega_e$ &
$E_i$ (Vm$^{-1}$) \\

\hline 

6.0 & 11 & 0.2 & 10.8 & 1.8 \\

5.0 & 11 & 0.3 & 10.7 & 2.3 \\

3.5 & 12 & 0.4 & 11.6 & 2.1 \\

3.25 & 12 & 0.3 & 11.7 & 1.4 \\

\hline

\end{tabular}

\vskip 0.5cm

Comparing Tables 1 and 2, we see that waves are excited 
with amplitudes exceeding $E_i$ in all
four cases. For the simulation with $u_{b\perp}=3.25v_{e0}$,  
$E/E_i \simeq 1.1$. This ratio rises to 1.2 for
$u_{b\perp} = 3.5v_{e0}$, 10 for $u_{b\perp} = 5v_{e0}$, and 19 for
$u_{b\perp} = 6v_{e0}$. In the latter two cases, as we have seen,
waves are excited with $\omega = \ell\Omega_e$, for which island
formation occurs regardless of the value of $E$. The fact that $E_m/E_i
\gg 1$ at higher values of $u_{b\perp}$ indicates that the phase
space in these simulations is characterized by strong
stochasticity. The waves rapidly collapse, however, soon after the
onset of strong electron acceleration.    
In the other two simulations, the peak amplitudes
are only just sufficient for island formation to occur, and it is
likely that little stochasticity occurs in the system phase space.     
The waves excited in these simulations decay more gradually than those
produced at higher $u_{b\perp}$. 

We now consider possible explanations for two of the results noted above:
the sharp rise in wave amplitude when the beam speed is raised from
$3.5v_{e0}$ to $5v_{e0}$; and wave collapse, which occurs in all four
simulations but is particularly rapid in the two simulations with
higher $u_{b\perp}$. 
As far as the dependence of wave amplitude on $u_{b\perp}$ is
concerned, the first point to note is that the unstable waves all
satisfy $\omega \simeq u_{b\perp}k$. In each simulation the total
number of computational particles is, of course, finite, 
the Maxwellian electron velocity distribution being initialized up to
$v_{\perp} \simeq 5v_{e0}$.
Thus, the beams with $u_{b\perp} = 5v_{e0}$ and $6v_{e0}$
excite waves with phase velocities exceeding the velocity of any
electron in the simulation: this is not so in the simulations with  
$u_{b\perp}=3.25v_{e0}$ and $3.5v_{e0}$. The minimum electron velocity  
required for wave--particle interactions is the phase
velocity of the wave: thus, only the slow beams
can excite waves capable of interacting with electrons at the start of
the simulations. The wave--particle interaction results in
electron acceleration, the energy for this being drawn from the
wave. This energy loss may account for the relatively low peak
electric field amplitudes of waves excited by the slow beams. 

The waves generated by the fast beams, on the other hand, cannot initially 
interact resonantly with the electron population, and so their amplitudes can
grow to levels much higher than $E_i$. Sufficiently high wave
amplitudes can activate a second acceleration mechanism, which arises
from particle trapping in the wave electric field (Karney
1978): electrons with an initially monotonic
decreasing distribution are re--distributed uniformly within the trap,
the result being a net increase in kinetic energy. The wave can trap
electrons with perpendicular velocities differing from the wave's
phase velocity by up to $v_{\rm tr}$, where 
$$v_{\rm tr} = \sqrt{\frac{eE}{mk}}. \eqno (15) $$
For $u_{b\perp} = 5v_{e0}$, the maximum electric field is 23$\,$Vm$^{-1}$
and the wavenumber $k$ is $5.7 \times 10^{-3} \times
2\pi\,$m$^{-1}$. In this case 
$v_{\rm tr} = 1.1 \times 10^7\,$ms$^{-1}\,\simeq 2.8v_{e0}$.
For $u_{b\perp} = 6v_{e0}$, the maximum field is 35$\,$Vm$^{-1}$
and $k = 4.8 \times 10^{-3} \times 2\pi\,$m$^{-1}$, so that
$v_{\rm tr} = 1.4 \times 10^7\,$ms$^{-1}\, \simeq 3.8v_{e0}$.
The waves excited in the simulations with higher beam speeds can thus
trap electrons deep within the electron thermal population: a large
number of electrons can then be pre--accelerated to velocities
comparable to the wave's phase velocity, with further acceleration
taking place via the stochastic mechanism discussed previously. A two--stage
process of this type 
was proposed by Karney (1978). Whereas the first burst of
wave activity in Fig. 3 contained more energy than the electron
population, the energy in the second burst was much lower than the
perpendicular electron kinetic energy by that stage of the simulation.
This may have been due to the first burst resulting in trapped electrons
populating the region of phase space at $v_{\perp} \simeq u_{b\perp}$,
via the trapping mechanism. The perpendicular electron velocity
distribution would then be considerably broader than it was initially,
with an effective thermal speed $v_e > v_{e0}$. The beam distribution,
on the other hand, did not change significantly during the simulation
(see Fig. 2), and so $u_{b\perp}/v_e < u_{b\perp}/v_{e0}$.    
The situation would then be similar to that of the simulations
with lower beam speeds, in which electrons can immediately absorb
energy from waves with $\omega \simeq ku_{b\perp}$, and one would expect 
any subsequent wave burst to have a peak energy much lower than that
of the electron population, as observed.

With regard to the second observation, wave collapse, it is
interesting to note that in every case the wave amplitude falls to a level well
below $E_i$: intuitively, one would have expected the waves to cease
interacting with electrons, and hence to reach a steady--state level, when
their amplitudes had fallen below $E_i$. The collapse may be
associated with changes in the dispersion characteristics of the wave
mode resulting from strong particle acceleration. Karney (1978)
justified his Hamiltonian approach by considering only stochastic
regions of phase space, at particle speeds (and hence wave phase
speeds) greatly exceeding $v_{e0}$. The stochastic regions thus lie in the 
high velocity tail of the initial Maxwellian electron distribution,
and most electrons are not initially affected by the wave--particle
interaction. However, in the simulations with $u_{b\perp} = 5
v_{e0}$, $6v_{e0}$ the reduction in $u_{b\perp}/v_e$ noted above
means that perpendicular electron speeds are no longer small compared
to the wave phase speed, and we find that there is a transition from the pure
Buneman instability shown in Fig. 8 to the more complicated
instability shown in Fig. 9: the latter, as we have discussed, has a
Buneman--like envelope, but also has cyclotronic features, and in fact
linear stability analysis shows that the variation of $\omega$ with $k$ in
this case is characteristic of the beam/electron Bernstein mode
discussed in Subsect. 3.1. As $u_{b\perp}/v_e$ falls, the maximum
growth rate drops considerably, but remains positive if 
the electrons retain a Maxwellian distribution. However, as we now
demonstrate, the electron distributions occurring in the simulations
are often far from Maxwellian. 

\subsection{Particle distributions}

From the simulation results we have evaluated the distribution
of perpendicular electron speeds $f(v_{\perp})$, defined such that 
$$ \int_0^{\infty}f(v_{\perp})dv_{\perp} = N_e\,, \eqno (16) $$
where $N_e$ is the total number of electrons in the simulation.   
With this definition, a Maxwellian velocity distribution is of the form 
$v_{\perp}e^{-v_{\perp}^2/2v_e^2}$, decreasing monotonically
for $v_{\perp} > v_e$. One advantage of plotting a distribution in this
way is that the thermal speed of a Maxwellian can be readily
identified graphically, being the speed at which
$df/dv_{\perp} = 0$. In Fig. 11 $f(v_{\perp})$ is plotted for 
$\tilde{t}=0$, $45$, $90$ and $135$ in the simulations with
$u_{b\perp} = 3.25v_{e0}$ (dash--dotted curves) and $u_{b\perp}
= 3.5v_{e0}$ (solid curves). The two curves are identical for
$\tilde{t}=0$, since the same initial electron temperature is used in
all four simulations. At
$\tilde{t} = 45$ the proton beams have generated hot electron tails,
peaking at $v_{\perp} \simeq 4v_{e0}$ ($u_{b\perp} = 3.25v_{e0}$) and 
$v_{\perp} \simeq 6v_{e0}$ ($u_{b\perp} = 3.5v_{e0}$). The maximum
electron speeds in the two cases are $7v_{e0}$ ($u_{b\perp} =
3.25v_{e0}$) and $10v_{e0}$ ($u_{b\perp} = 3.5v_{e0}$). At $\tilde{t}
= 90$ the slower beam has produced a local maximum in $f(v_{\perp})$ 
at $5v_{e0}$, and a high velocity cutoff at $10v_{e0}$. The local
maximum has become less pronounced at $\tilde{t} = 135$.
In the case of $u_{b\perp} = 3.5v_{e0}$ a local maximum can be seen at 
$\tilde{t} = 45$: by $\tilde{t} = 90$, however, the distribution is
monotonic decreasing above a speed only slighly higher than the
initial thermal speed. By the end of this simulation
$f(v_{\perp})$ extends up to 12$v_{e0}$.

In Fig. 12 $f(v_{\perp})$ is shown for $\tilde{t}=0$, $20$, $40$ and
$70$ in the simulations with $u_{b\perp} = 5v_{e0}$ (dash--dotted
curves) and $u_{b\perp} = 6v_{e0}$ (solid curves). At $\tilde{t}=20$
the two distributions have local maxima at $v_{\perp} \gg v_{e0}$,
as in the second frame of Fig. 11: for $u_{b\perp} = 5v_{e0}$ the
distribution peaks locally at $v_{\perp} \simeq 10v_{e0}$ and 
falls to zero at $v_{\perp} \simeq 18v_{e0}$. The corresponding
figures at the same stage of the simulation with $u_{b\perp} =
6v_{e0}$ are $12v_{e0}$ and $22v_{e0}$. By $\tilde{t}=40$, the local
maxima still exist and, indeed, the bumps--on--tail containing these
maxima actually comprise most of the electron population in both
cases. By this time the high velocity tails extend to 
$v_{\perp} \simeq 25 - 30v_{e0}$. Local maxima close to the original thermal
speeds $v_{e0}$ still exist, but these have disappeared by $\tilde{t}
= 70$. The strong wave bursts in these simulations  
occur before $\tilde{t}=20$: after this time, a weaker, more broadband
instability occurs at lower $\tilde{k}$, but still with $\tilde{\omega}
\simeq \tilde{k}u_{b\perp}/v_{e0}$. To model this instability,
we can approximate the solid curve at $\tilde{t}=20$ in Fig. 12 by
superposing two Maxwellians, with thermal velocities $v_{ec}=v_{e0}$,
$v_{eh} = 10v_{e0}$ and densities $n_{ec} = 0.38n_e$,
$n_{eh} = 0.62n_e$. The solid curve in Fig. 13 shows the true distribution 
at $\tilde{t}=20$; the
dashed curve shows the bi--Maxwellian fit to this distribution. The
match is not exact, but is close enough to suggest that we can
model wave excitation at this stage of the simulation  by
solving a modified version of the dispersion relation
[Eq. (2)], with the parameters of the dashed curve defining the
electron distribution. Results indicate an electron Bernstein
instability with maximum growth rate 
$\gamma \simeq 0.08\Omega_e$ at $\tilde{k} \simeq 1.4$,
$\tilde{\omega} \simeq 8.2\Omega_e$: the wavenumbers are 
consistent with those of fluctuations appearing at
$\tilde{t} \ge 20$ in Fig. 5. 

The electron distributions at $\tilde{t} = 70$ in the simulations with
$u_{b\perp}=5v_{e0}$ and $u_{b\perp} = 6v_{e0}$ (Fig. 12) can both be
approximated by single Maxwellians, respectively with $v_e \simeq
8v_{e0}$ and $v_e \simeq 12v_{e0}$. The proton beam--excited Buneman
instability can thus produce electron distributions whose
perpendicular thermal speeds exceed the velocities of the ion
beams which produced them. The fastest--moving electrons have
$v_{\perp} \gg u_{b\perp}$. This phenomenon, observed in all four
simulations, is further strong evidence for nonlinear processes
playing a role in electron acceleration: in the quasi--linear limit,
unmagnetized electrons of a particular speed $v_0$ can only interact with waves
whose phase speed equals $v_0$, and the range of wave phase speeds is
determined in turn by the ion beam speeds. In the case of $u_{b\perp}
= 6v_{e0}$, the final electron temperature is about 11.5$\,$keV: this
is easily sufficient to account for thermal X--ray emission observed
from SNRs such as Cas A (Papadopoulos 1988). Individual   
electron energies up to several tens of keV were obtained in the
simulations.   

\section{Conclusions and Discussion}

Using particle--in--cell (PIC) simulations and linear stability theory, we have
shown that electrostatic waves in the electron plasma range, excited
by ions reflected from a high Mach number perpendicular shock, can
effectively channel the free energy of the shock into
electrons. Such shocks are known to be associated with SNRs, and the
processes investigated in this paper may thus help to account for both
X--ray thermal bremsstrahlung and the creation of ``seed'' electron
populations for diffusive shock acceleration to MeV and GeV energies
in such objects. The simulation results provide confirmation of a proposal by
Papadopoulos (1978) and Cargill \& Papadopoulos (1978)
that streaming between reflected ions and upstream electrons can give
rise to a strong Buneman instability. Whereas these authors
assumed that the sole effect of the Buneman instability would be electron
heating, the PIC simulations show acceleration -- the development of strongly
non--Maxwellian, anisotropic features in electron velocity
distributions. The maximum electron velocities are considerably higher
than those expected on the basis of quasi--linear theory: this implies
that nonlinear wave--particle interactions are contributing to the
acceleration. When the beam speed is greater than about four times the
initial electron thermal speed, thermalization of the electron
population is observed after saturation of the Buneman instability,
the final electron temperature being of the order of 10--12$\,$keV.     

It is possible that the acceleration process identified in this paper
may be relevant to oblique shocks as well as strictly perpendicular 
ones. A necessary requirement is the presence of 
reflected ion beams, which have been observed (Sckopke et al. 1983) 
upstream of both quasi--parallel and quasi--perpendicular regions of
the Earth's bow shock (the term ``quasi--perpendicular'' is conventionally
used to describe a shock at which the angle $\theta_{Bn}$ between the shock
normal and the upstream magnetic field is greater than 45$^{\circ}$). 
Leroy et al. (1982) used a hybrid code to simulate ion reflection at shocks 
with a range of values of $\theta_{Bn}$, finding very similar results for  
$\theta_{Bn} = 80^{\circ}$ and $\theta_{Bn} = 90^{\circ}$. They inferred from 
this that hybrid simulations which use strictly perpendicular geometry may be
compared with observational data of quasi--perpendicular shocks.
Additional necessary requirements 
for electron acceleration via the Buneman instability are that the 
projection of the reflected ion beam velocity onto the plane perpendicular to 
the upstream field exceed the upstream electron thermal speed, and that the
plasma be weakly magnetized, in the sense that $\omega_{pe} > \Omega_e$. When 
these conditions are satisfied locally, the Buneman instability
will occur. Whether this instability is sufficient to produce significant
electron energization at oblique shocks as well as perpendicular and nearly
perpendicular ones remains to be demonstrated, however. In the simulations 
presented in this paper, acceleration occurred on timescales shorter than an 
ion cyclotron period. It was not necessary to represent the shock foot 
structure, since this has dimensions of the order of a reflected ion Larmor 
radius, and for this reason $\theta_{Bn}$ is not explicitly a parameter in our 
model. Leroy et al. (1982) have noted, however, that extrapolations of results
obtained for nearly perpendicular shocks ($80^{\circ} \le \theta_{Bn}
\le 90^{\circ}$) to more oblique shocks should be
treated with caution, since the physical processes governing the shock
structure can be expected to change as $\theta_{Bn}$ is reduced.

The simulation results and our analysis of them suggest that the
damping of waves in the electron plasma range
at perpendicular SNR shocks could 
provide a solution to the cosmic ray electron injection problem. Although 
wave--particle interactions at such shocks have been invoked 
previously in this context (Galeev 1984; Papadopoulos 
1988; Cargill \& Papadopoulos 1988; Galeev et al. 
1995; McClements et al. 1997), the simulation
results presented here contain several new features. These include:
the acceleration, rather than heating, of electrons via the Buneman
instability; the acceleration of electrons to speeds exceeding those
of the shock--reflected ions producing the instability; and strong acceleration
of electrons perpendicular to the magnetic field. The wave--particle  
mechanisms proposed by Galeev (1984) and McClements et
al. (1997) gave rise to electron acceleration primarily
along the magnetic field. Diffusive shock
acceleration, which is probably essential for the production of 
ultra--relativistic electrons, can occur when the electrons
have magnetic rigidities comparable to those of ions flowing
into the shock. Since, however, the diffusive shock mechanism requires 
electrons to be rapidly scattered, its efficacy 
does not depend sensitively on the initial pitch angle distribution.
The geometry of the simulations described here ({\bf B} 
perpendicular to the one space dimension) excludes the 
possibility of acceleration by electrostatic waves along the parallel 
direction. The present model is complementary to those of Galeev (1984), 
Galeev et al. (1995) and McClements et al. (1997), in that it provides an 
alternative means of energizing electrons at perpendicular shocks. At a 
real SNR shock, perpendicular acceleration via the Buneman 
instability and parallel acceleration via wave excitation at $\omega < 
\Omega_e$ are both likely to occur. PIC simulations in two space 
dimensions would make it possible to assess quantitatively the 
relative contributions of these two types of instability to electron 
energization under a range of conditions.   
 
We discuss finally our neglect of the finite plasma current present in
the foot region of perpendicular shocks. This approximation does not
appear to have introduced unrealistic elements into our simulation
results, except insofar as the absence of a finite drift between the
electrons and background protons removes a possible source of drive
for the ion acoustic instability, invoked as one of the electron
heating mechanisms in the model of Papadopoulos
(1988). However, if the background protons and electrons flowing
into the shock foot are approximately isothermal, it seems
unlikely that any instability excited by their relative streaming
could result in a significant transfer of energy from one species
to the other. Another possibility is that ion acoustic instability
could result from the streaming of beam protons relative to
electrons. This would require, however, the electron temperature to be
extremely large compared to the beam proton temperature (ion Landau
damping strongly suppresses the instability when the temperatures are
equal): even in the
simulation which produced an effective electron temperature of 80
times the initial temperature, this condition was not satisfied. Thus,
the simulation parameters were such that the ion acoustic instability
could not occur. It is interesting to note, however, that the curve
corresponding to $u_{b\perp} = 6v_{e0}$ in the lower frame of Fig. 1 bears
a certain resemblance to a curve in Fig. 5 of Papadopoulos' 1988
paper (Fig. 5), showing schematically the predicted variation of
electron temperature with distance in
a quasi--perpendicular shock foot: in both cases, there is a very rapid rise
in the total electron energy, resulting from strong Buneman instability,
followed by a more gradual rise, which coincides in the simulations
with the excitation of electron
Bernstein modes. The latter may play a role in the simulations which is  
similar to that of the ion acoustic mode in the model of Papadopoulos.  

\vskip 0.5cm

\noindent {\bf Acknowledgements.} This work was supported by the 
Commission of the European Communitities, under TMR Network Contract 
ERB--CHRXCT98-- 0168, by the UK Department of Trade and Industry, and by
EURATOM. S. C. Chapman was supported by a PPARC lecturer fellowship.

\vskip 1.0cm

\centerline{REFERENCES}

\vskip 0.5cm

\ref{Achterberg A., Blandford R.D., Reynolds S.P., 1994, A\&A 281, 220}

\ref{Anderson K.A., Lin R.P., Martel F., Lin C.S., Parks G.K., Reme H., 1979, 
Geophys. Res. Lett. 6, 401}

\ref{Axford W.I., Leer E., Skadron G., 1977, in: Proceedings of the 15th 
International Cosmic Ray Conference, Christov C.Y. (ed.), Central Research 
Institute for Physics, Budapest, vol. 11, 132} 

\ref{Bell A.R., 1978, MNRAS 182, 147} 

\ref{Bessho N., Ohsawa Y., 1999, Phys. Plasmas 6, 3076}
                 
\ref{Blandford R.D., Ostriker J.P., 1978, ApJ 221, L29} 

\ref{Buneman O., 1958, Phys. Rev. Lett. 1, 8} 

\ref{Cargill P.J., Papadopoulos K., 1988, ApJ 329, L29} 

\ref{Denavit J., Kruer W.L., 1980, Comments Plasma Phys. Cont. Fusion 6, 35}

\ref{Devine P., 1995, Ph.D. Thesis, University of Sussex}

\ref{Galeev A.A., 1984, Sov. Phys. JETP 59, 965}

\ref{Galeev A.A., Malkov M.A., V\"olk H.J., 1995, J. Plasma Phys. 54, 59}

\ref{Karney C.F.F., 1978, Phys. Fluids 21, 1584} 

\ref{Kirk J.G., Heavens A.F., 1989, MNRAS 239, 995} 

\ref{Koyama K., Petre R., Gotthelf E.V, Hwang, U., Matsuura M., Ozaki M., Holt 
S.S., 1995, Nature 378 255}

\ref{Krymsky G.F., 1977, Dokl. Akad. Nauk. SSSR 234, 1306}

\ref{Laming J.M., 1998, ApJ 499, 309}

\ref{Leroy M.M., Winske D., Goodrich C.C., Wu C.S., Papadopoulos, K., 1982, J. 
Geophys. Res. 87, 5081}

\ref{Levinson A., 1996, MNRAS 278, 1018}

\ref{McClements K.G., Bingham R., Su J.J., Dawson J.M., Spicer D.S., 1993, 
ApJ 409, 465}

\ref{McClements K.G., Dendy R.O., Bingham R., Kirk J.G., Drury L.O'C., 1997, 
MNRAS 291, 241}

\ref{Melrose D.B., 1986, Instabilities in Space and Laboratory Plasmas, 
Cambridge University Press}

\ref{Papadopoulos K., 1988, Ap\&SS 144, 535}

\ref{Pohl M., Esposito J.A., 1998, ApJ 507, 327}

\ref{Quest K.B., 1986, J. Geophys. Res. 91, 8805}

\ref{Sckopke N., Paschmann G., Bame S.J., Gosling J.T., Russell C.T., 1983, J. 
Geophys. Res. 88, 6121} 

\ref{Willingale R., West R.G., Pye J.P., Stewart G.C., 1996 MNRAS 278, 479}

\ref{Woods L.C., 1969, J. Plasma Phys. 3, 435}

\newpage

\begin{figure}
\setlength{\unitlength}{1cm}
\begin{picture}(0.0,9.0)
\put(0.3,2.0){\includegraphics{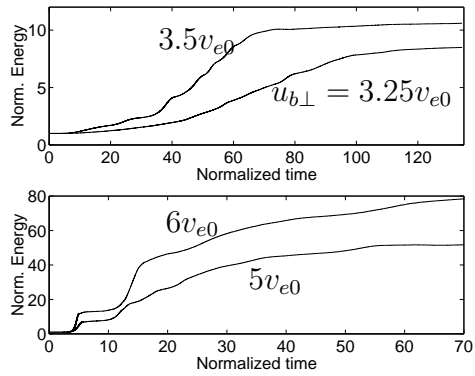}}
\put(4.4,7.9){$u_{b\perp} = 3.25v_{e0}$}
\put(2.9,8.6){$3.5v_{e0}$}
\put(4.1,5.4){$5v_{e0}$}
\put(3.0,6.17){$6v_{e0}$}
\end{picture}
\parbox[b]{8.6cm}{\caption[]{Total electron perpendicular kinetic
energy, normalized to its initial value, versus simulation time in
electron cyclotron periods $2\pi/\Omega_e$, for several values of
$u_{b\perp}/v_{e0}$. Energy transfer to electrons is much more rapid at
$u_{b\perp} = 5v_{e0}$ and $u_{b\perp} = 6v_{e0}$ than it is at lower
beam speeds.}}%
\label{energy1}
\end{figure}

\begin{figure}
\setlength{\unitlength}{1cm}
\begin{picture}(0.0,10.0)
\put(0.3,2.0){\includegraphics{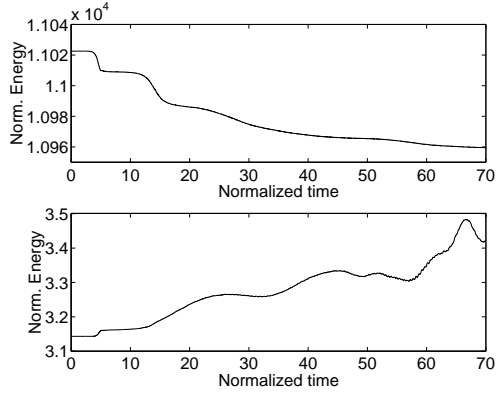}}
\end{picture}
\parbox[b]{8.6cm}{\caption[]{Normalized beam proton (upper plot) and
background proton (lower plot) perpendicular kinetic energies (both
normalized to initial electron thermal energy) versus simulation time for the
simulation with $u_{b\perp} = 6v_{e0}$. The beam proton energy drops
by less than 1\%; the background proton energy increases by
approximately 10\%.}}%
\label{energy2}
\end{figure}

\begin{figure}
\setlength{\unitlength}{1cm}
\begin{picture}(0.0,8.0)
\put(0.3,0.5){\includegraphics{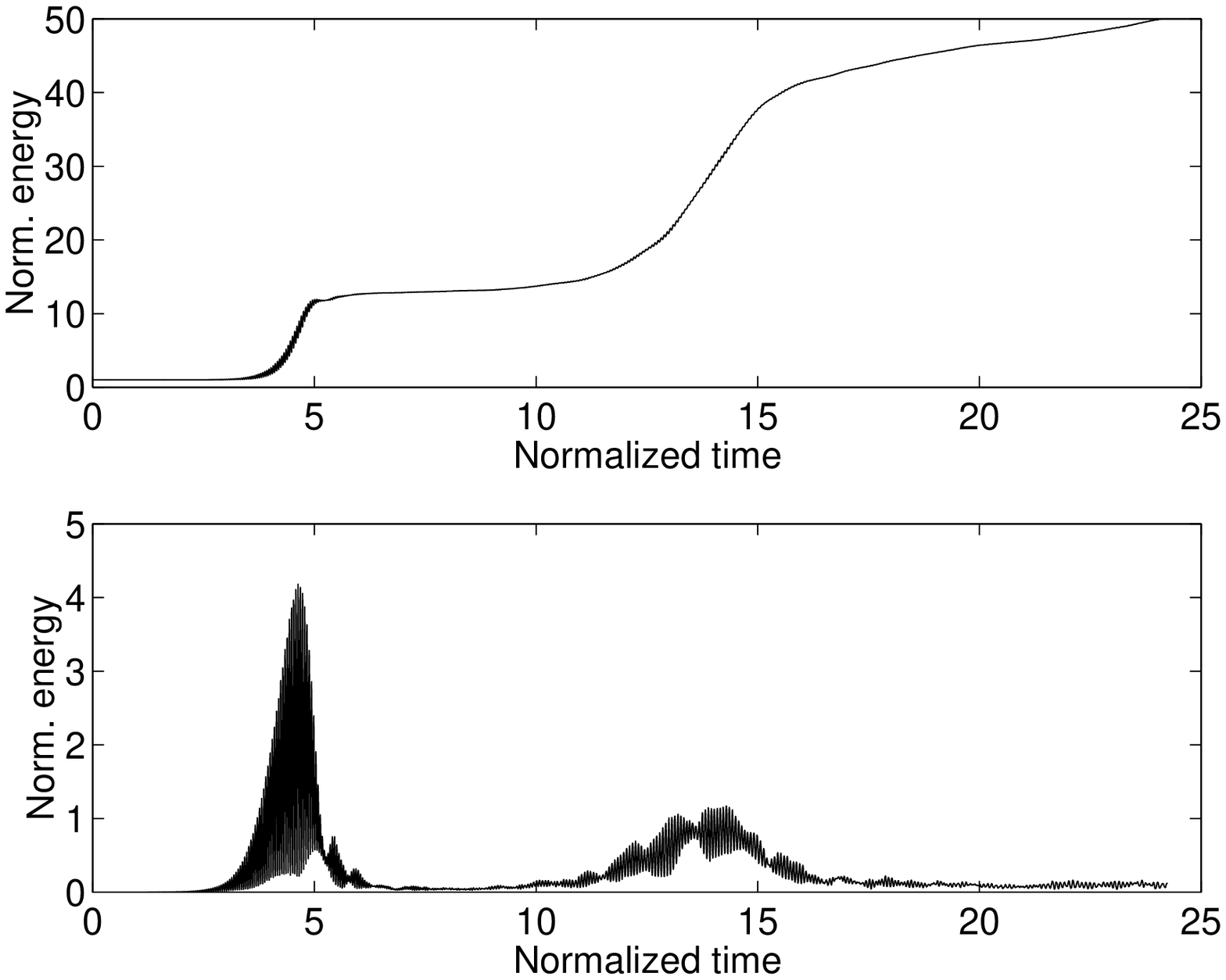}}
\end{picture}
\parbox[b]{8.6cm}{\caption[]{Time evolution of perpendicular electron
kinetic energy (upper plot) and electrostatic field energy (lower
plot) in the simulation with $u_{b\perp} = 6v_{e0}$.}}%
\label{waves}
\end{figure}

\begin{figure}
\setlength{\unitlength}{1cm}
\begin{picture}(0.0,8.0)
\put(0.3,0.0){\includegraphics{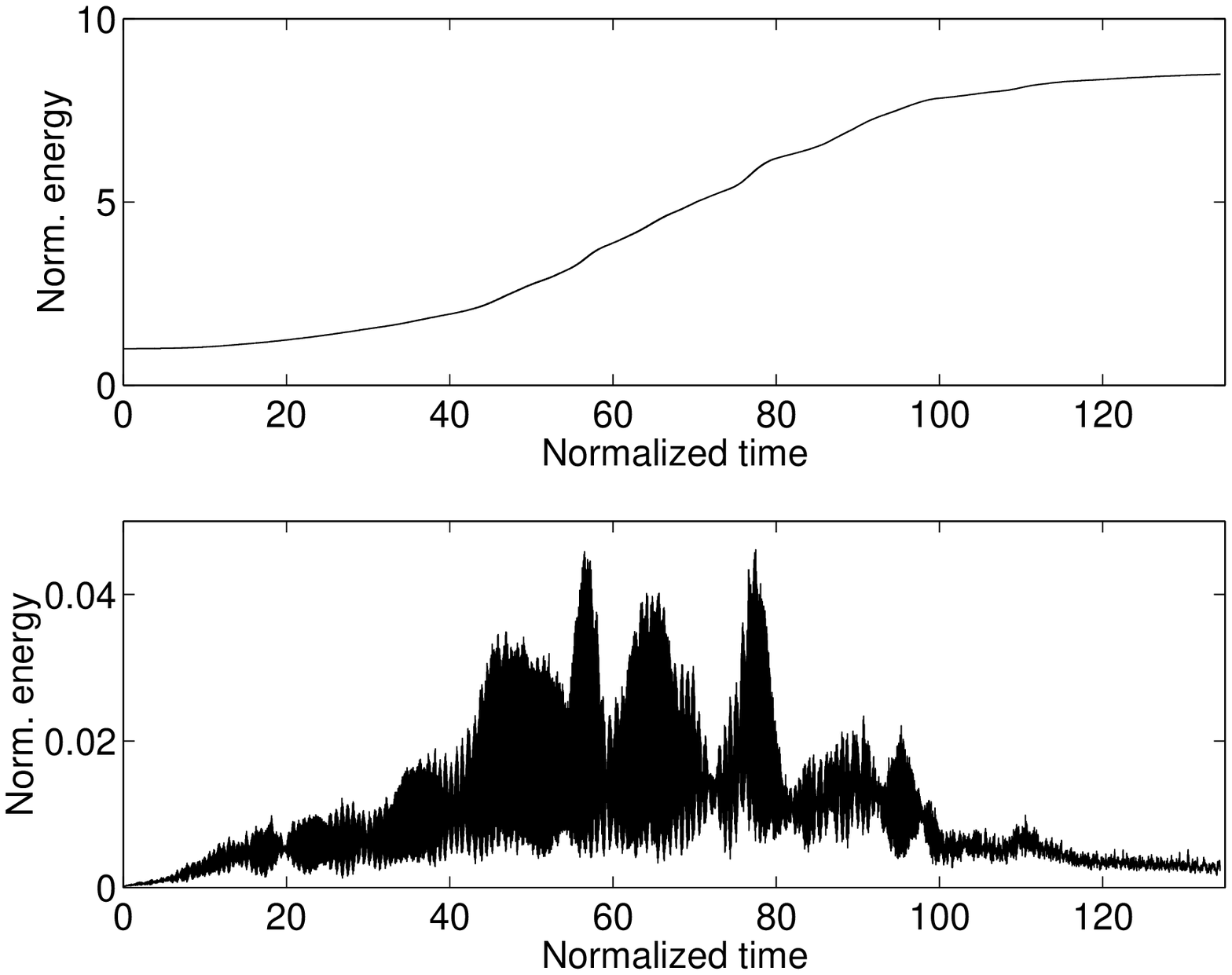}}
\end{picture}
\parbox[b]{8.6cm}{\caption[]{Time evolution of perpendicular electron
kinetic energy (upper plot) and electrostatic field energy (lower
plot) in the simulation with $u_{b\perp} = 3.25v_{e0}$.}}%
\label{dispersion}
\end{figure}

\begin{figure}
\setlength{\unitlength}{1cm}
\begin{picture}(0.0,8.0)
\put(0.3,0.0){\includegraphics{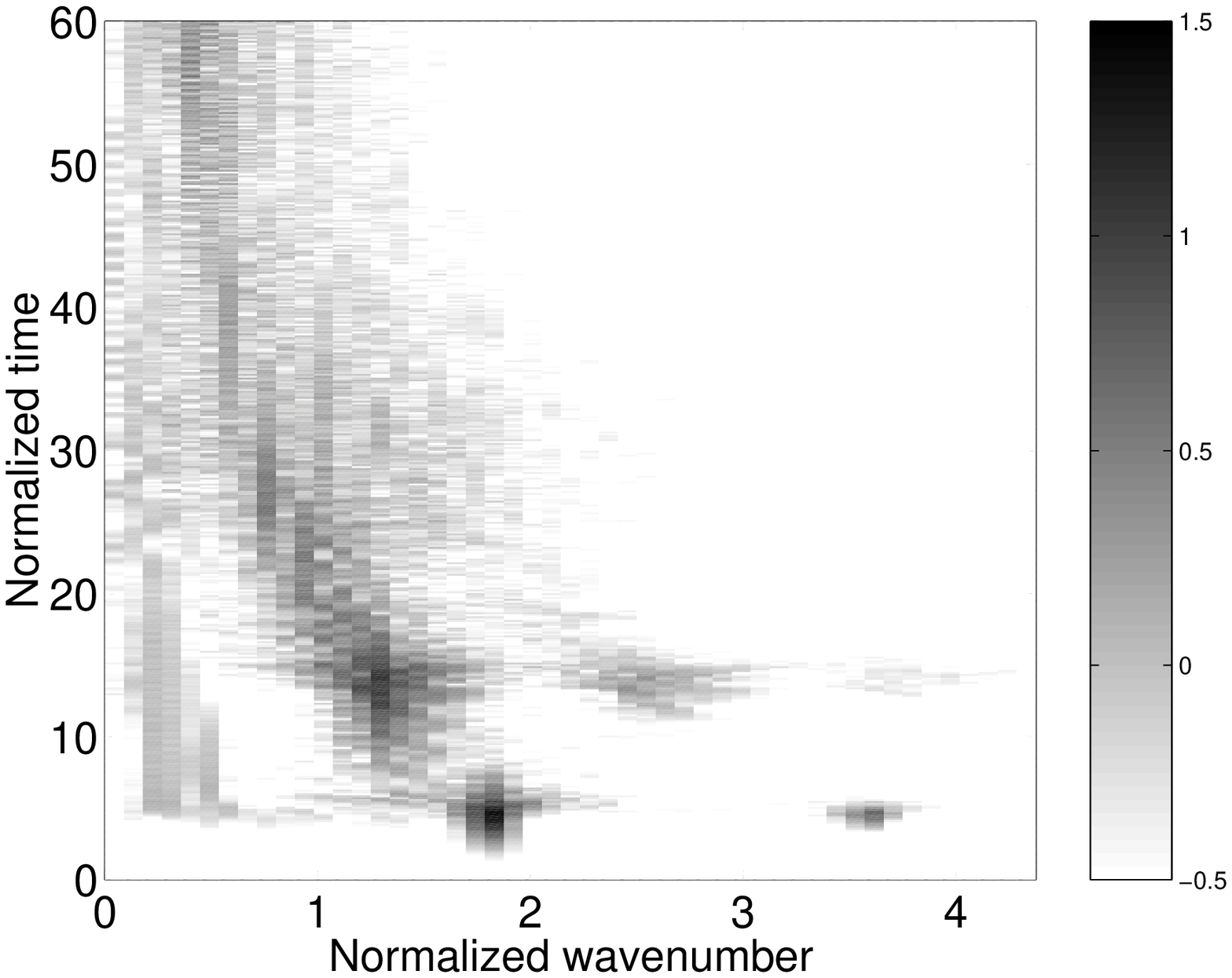}}
\end{picture}
\parbox[b]{8.6cm}{\caption[]{Base 10 logarithm of electric field
amplitude (Vm$^{-1}$) versus $\tilde{k}$ and $\tilde{t}$
in the simulation with $u_{b\perp} = 6v_{e0}$.}}%
\label{phasespace}
\end{figure}

\begin{figure}
\setlength{\unitlength}{1cm}
\begin{picture}(0.0,8.0)
\put(0.3,0.0){\includegraphics{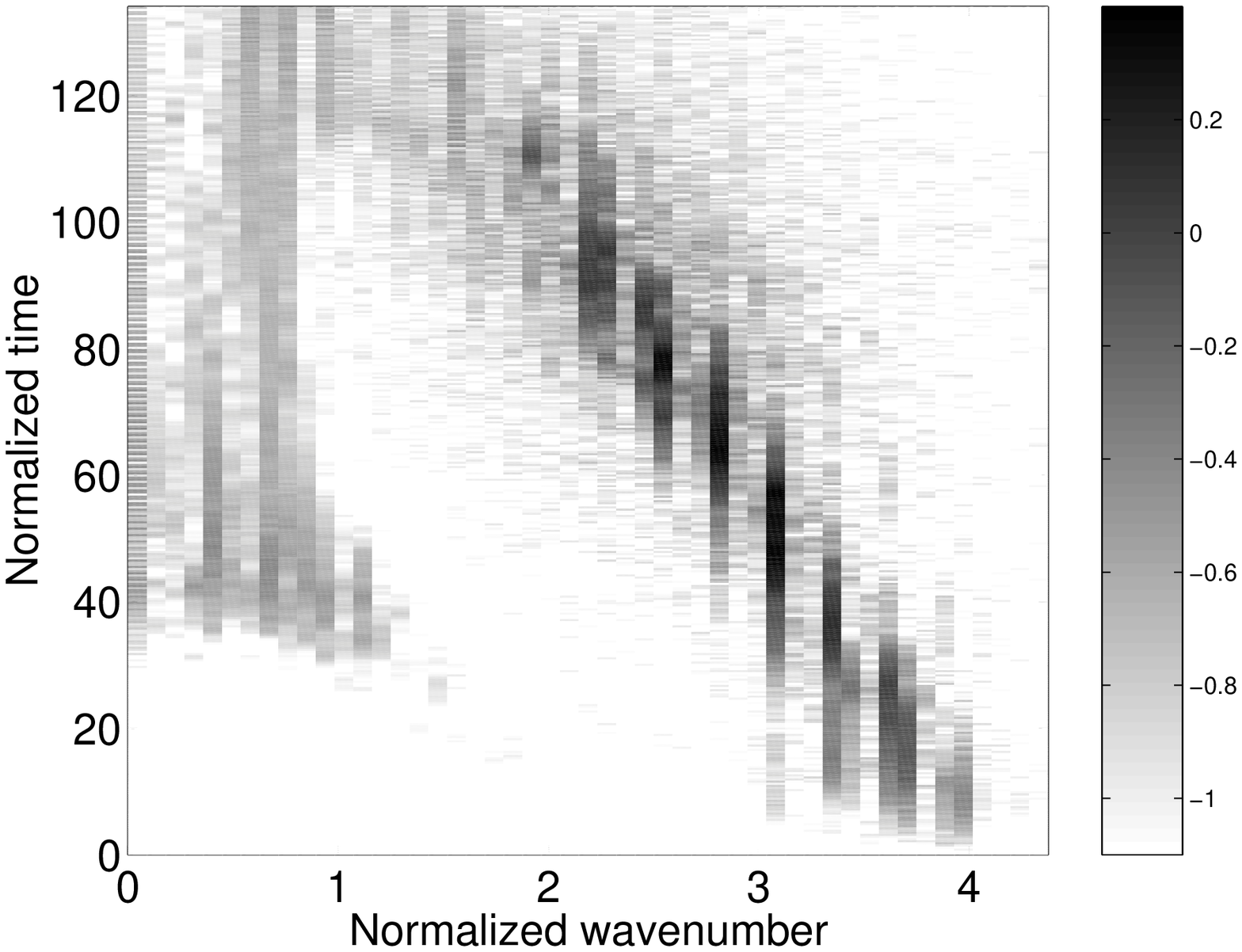}}
\end{picture}
\parbox[b]{8.6cm}{\caption[]{Base 10 logarithm of electric field
amplitude (Vm$^{-1}$) versus $\tilde{k}$ and $\tilde{t}$
in the simulation with $u_{b\perp} = 3.25v_{e0}$.}}%
\label{distribution}
\end{figure}

\begin{figure}
\setlength{\unitlength}{1cm}
\begin{picture}(0.0,11.7)
\put(-0.7,1.0){\includegraphics{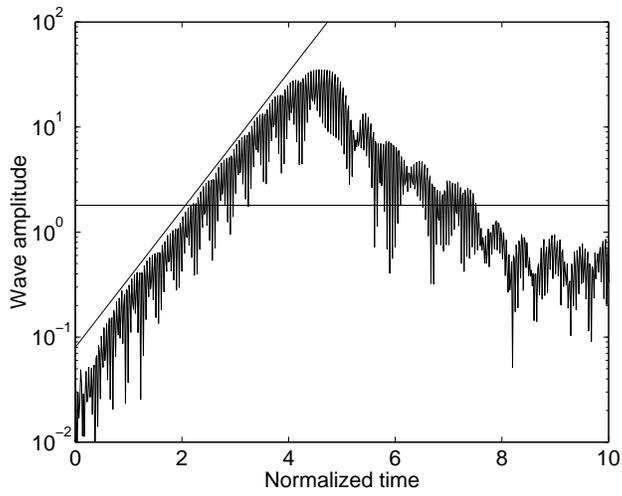}}
\end{picture}
\parbox[b]{8.6cm}{\caption[]{Time evolution of electric field
amplitude (Vm$^{-1}$) at $\tilde{k}=1.8$ after the onset of
instability in the simulation with $u_{b\perp} = 6v_{e0}$. The horizontal
line is $E_i$ for this wave; the diagonal line shows an exponential fit to the
growth rate with $\gamma/\Omega_e = 0.24$.}}
\label{amplitude6.0}
\end{figure}

\begin{figure}
\setlength{\unitlength}{1cm}
\begin{picture}(0.0,10.0)
\put(0.1,9.5){\includegraphics{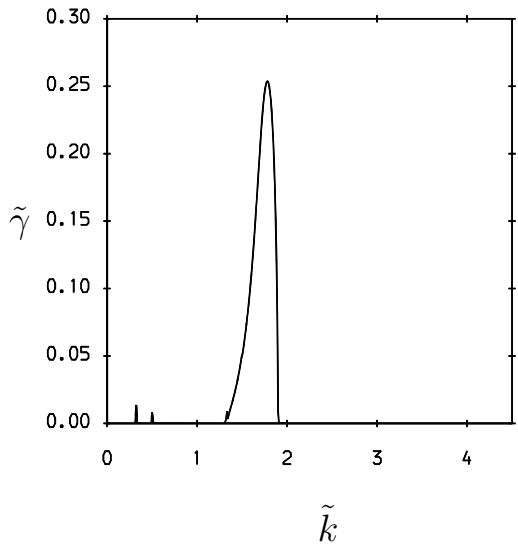}}
\put(0.4,6.3){\large $\tilde{\gamma}$}
\put(4.5,2.2){\large $\tilde{k}$}
\end{picture}
\parbox[b]{8.6cm}{\caption[]{Predicted linear growth rates of waves
with $\omega > \Omega_e$ when the beam
speed is $6v_{e0}$. The other dispersion relation parameters
correspond to the initial conditions of all four simulations.}} 
\label{hotbeam1}
\end{figure}

\begin{figure}
\setlength{\unitlength}{1cm}
\begin{picture}(0.0,10.0)
\put(0.1,9.0){\includegraphics{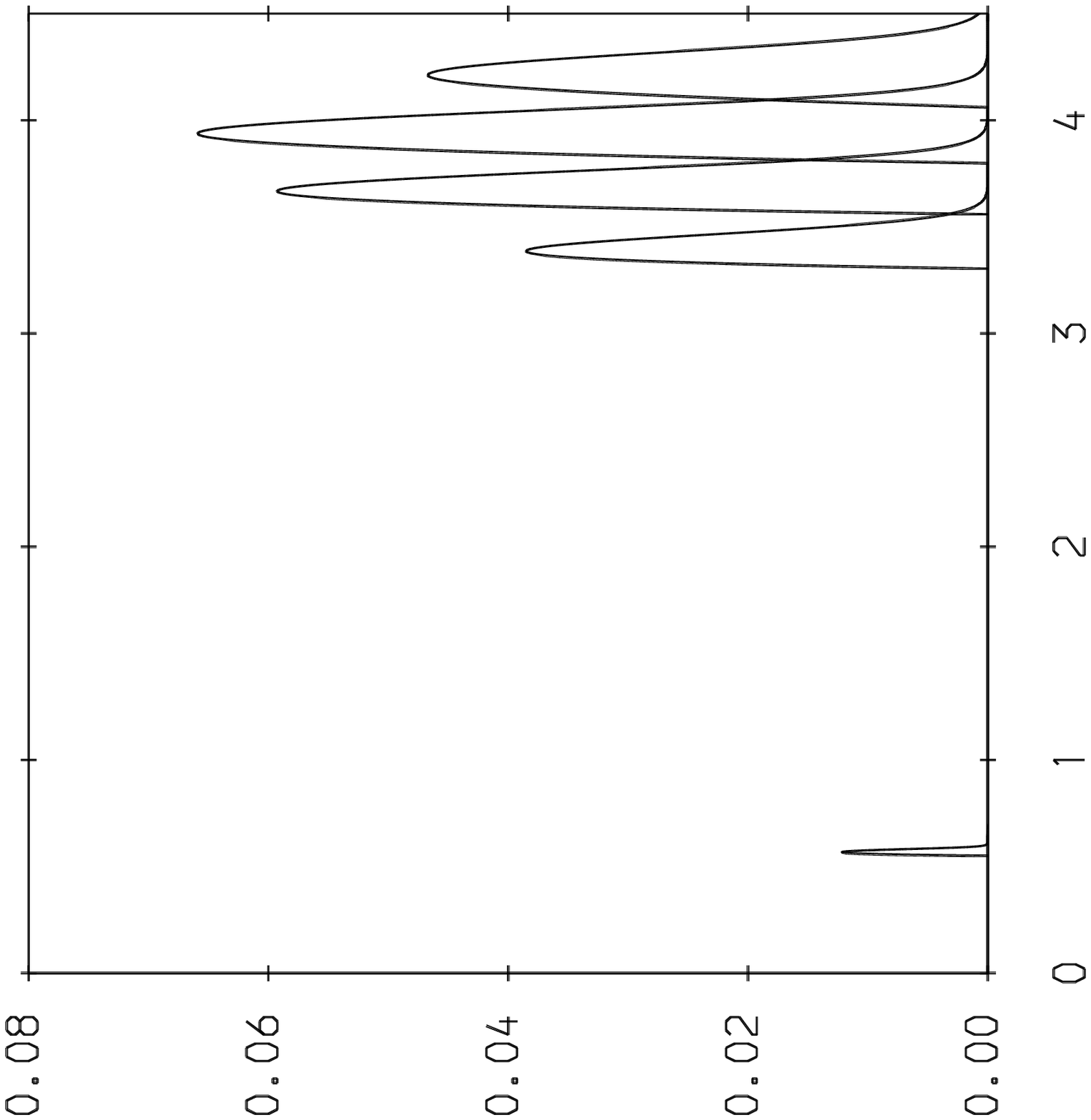}}
\put(0.4,6.8){\large $\tilde{\gamma}$}
\put(4.5,2.2){\large $\tilde{k}$}
\end{picture}
\parbox[b]{8.6cm}{\caption[]{Predicted linear growth rates of waves
with $\omega > \Omega_e$ when the beam speed is $3.25v_{e0}$.}}
\label{hotbeam2}
\end{figure}

\begin{figure}
\setlength{\unitlength}{1cm}
\begin{picture}(0.0,11.7)
\put(-0.7,1.0){\includegraphics{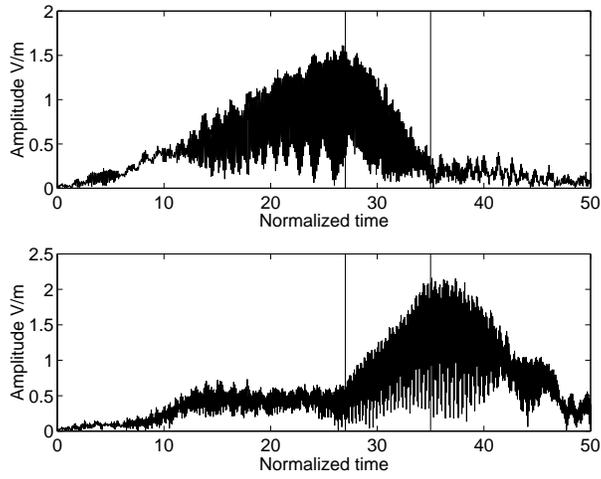}}
\end{picture}
\parbox[b]{8.6cm}{\caption[]{Time evolution of electric field
amplitude (Vm$^{-1}$) at $\tilde{k} = 3.6$ (upper plot) and $\tilde{k}
= 3.3$ (lower plot) after the onset of instability in the simulation
with $u_{b\perp} = 3.25v_{e0}$. The vertical lines indicate a period
in which there is an anti--correlation between the two wave
amplitudes.}} 
\label{amplitude3.25}
\end{figure}

\begin{figure}
\setlength{\unitlength}{1cm}
\begin{picture}(0.0,11.7)
\put(-0.7,0.0){\includegraphics{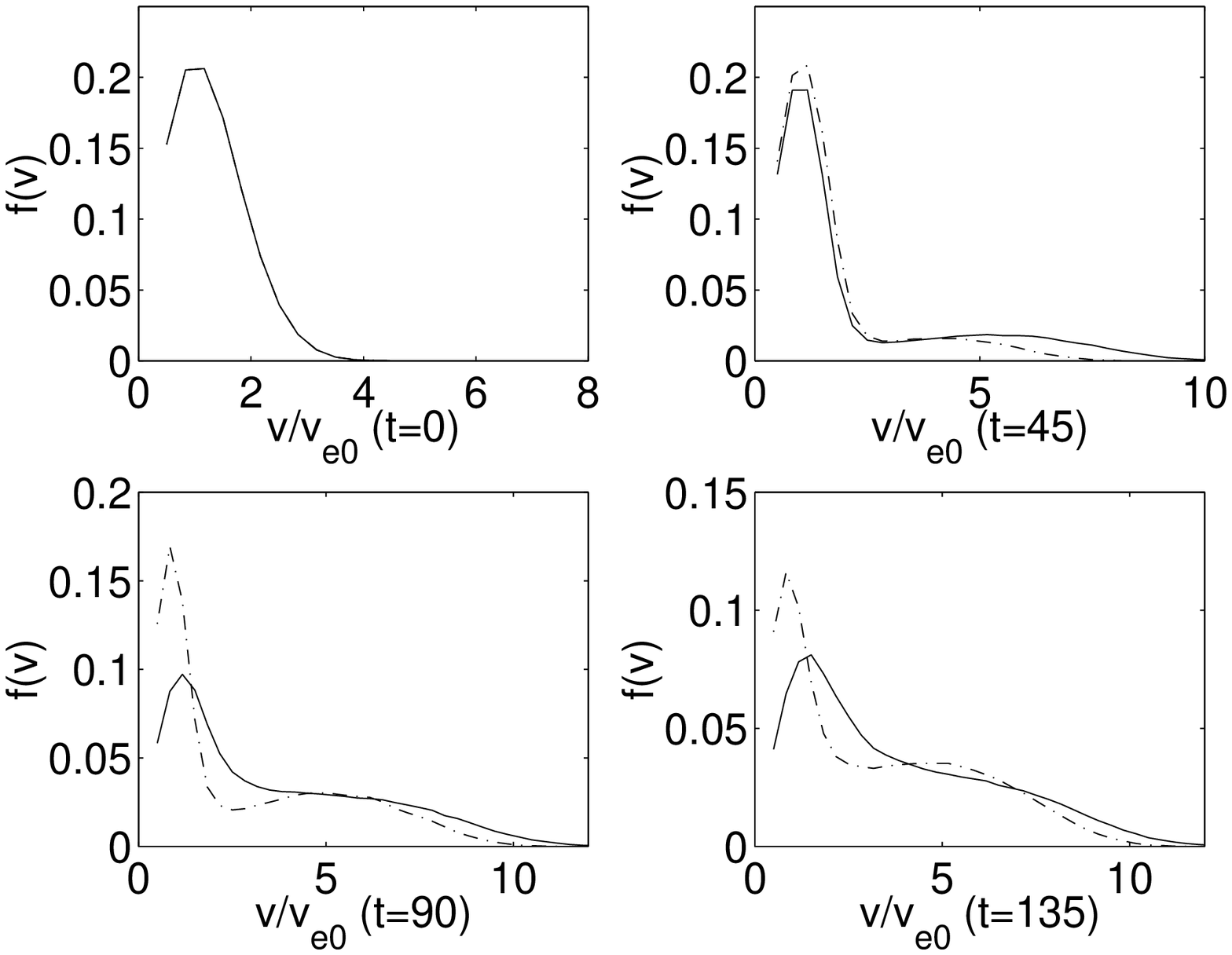}}
\end{picture}
\parbox[b]{8.6cm}{\caption[]{Normalized perpendicular electron
speed distributions at various times in the simulations with $u_{b\perp} =
3.25v_{e0}$ (dash--dotted lines) and $u_{b\perp} = 3.5v_{e0}$ (solid
lines).}}
\label{phase1}
\end{figure}

\begin{figure}
\setlength{\unitlength}{1cm}
\begin{picture}(0.0,10.7)
\put(-0.7,0.0){\includegraphics{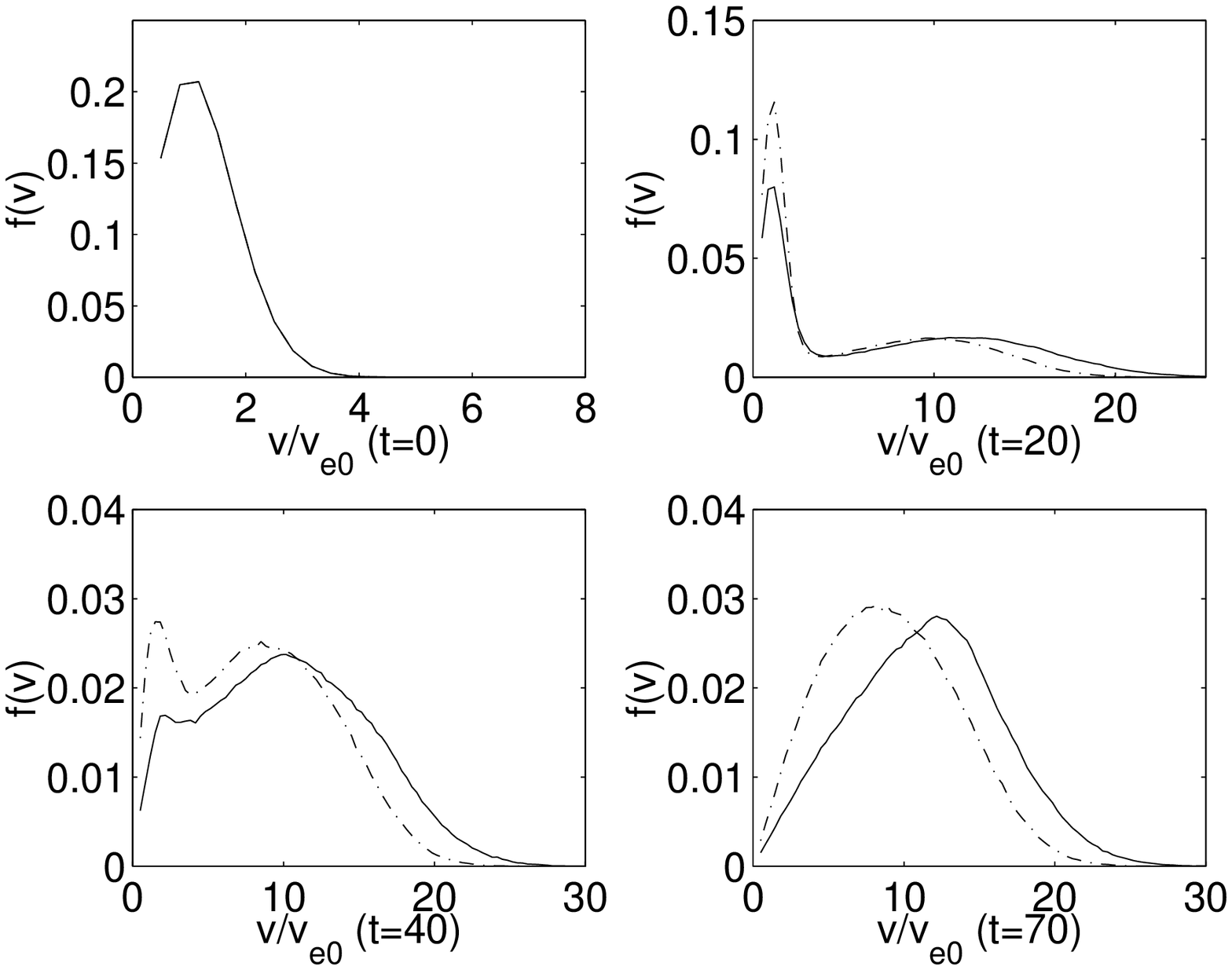}}
\end{picture}
\parbox[b]{8.6cm}{\caption[]{Normalized perpendicular electron
speed distributions at various times in the simulations with $u_{b\perp} =
5v_{e0}$ (dash--dotted lines) and $u_{b\perp} = 6v_{e0}$ (solid
lines).}}
\label{phase2}
\end{figure}

\begin{figure}
\setlength{\unitlength}{1cm}
\begin{picture}(0.0,10.7)
\put(0.0,1.0){\includegraphics{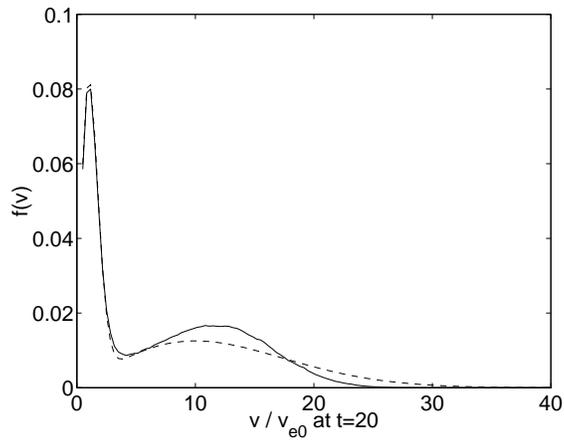}}
\end{picture}
\parbox[b]{8.6cm}{\caption[]{Normalized perpendicular electron
speed distribution at $\tilde{t}=20$ in the simulation with $u_{b\perp}
=6{v}_{e0}$. The solid curve is the distribution sampled in the
simulation; the dashed curve shows a bi--Maxwellian fit.}}
\label{fit}
\end{figure}

\end{document}